%% file: bare_jrnl_arxiv.tex
\documentclass[journal]{IEEEtran}
\ifCLASSINFOpdf
\else
\fi
\usepackage{amsmath,amssymb,amsfonts}   
\usepackage{bm}                         
\usepackage{graphicx}                   
\usepackage{epstopdf}
\usepackage{psfrag}                     
\usepackage{tikz}
\usepackage{pgfplots}
\usepackage{subcaption}
\usepackage{amsmath,graphicx}
\usepackage[ruled]{algorithm2e}
\usepackage{multirow}
\usepackage{color}
\input macros.tex

\begin{document}
%
\title{Using \cca{Double} Contractions to Derive the Structure of Slice-Wise Multiplications of Tensors with Application\cca{s} to Semi-Blind MIMO OFDM}
%
%
%

\author{Kristina~Naskovska, Andr\'{e}~L.~F.~de~Almeida, Martin~Haardt 
\thanks{K.~Naskovska and M.~Haardt are with the Communications Research Laboratory,
Ilmenau University of Technology, P. O. Box 100565, D-98684 Ilmenau, Germany.}
\thanks{A.~de~Almeida is with the Department of Teleinformatics Engineering, Federal University of Cear\'{a} (UFC), Fortaleza, Brazil.}
\thanks{Manuscript received Month Day, Year; revised Month Day, Year.}}

\maketitle

\begin{abstract}
The slice-wise multiplication of two tensors is required in a variety of tensor decompositions (including PARAFAC2 and PARATUCK2) and is encountered in many applications, including the analysis of multidimensional biomedical data (EEG, MEG, etc.) or multi-carrier MIMO systems. In this paper, we propose a new tensor representation that is not based on a slice-wise (matrix) description, but can be represented by a double contraction of two tensors. Such a double contraction of two tensors can be efficiently calculated via generalized unfoldings. It leads to new tensor models of the investigated system that do not depend on the chosen unfolding and reveal the tensor structure of the data model (such that all possible unfoldings can be seen at the same time). As an example, we apply this new concept to the design of new receivers for multi-carrier MIMO systems in wireless communications. In particular, we consider MIMO OFDM systems with and without Khatri-Rao coding. The proposed receivers exploit the channel correlation between adjacent subcarriers, require the same amount of training symbols as traditional OFDM techniques, but have an improved performance in terms of the symbol error rate. Furthermore, we show that the spectral efficiency of the Khatri-Rao coded MIMO-OFDM can be increased by introducing ''random coding" such that the ''coding matrix" also contains useful information symbols. Considering this transmission technique, we derive a tensor model and two types of receivers for randomly coded MIMO-OFDM systems using the double contraction of two tensors.
\end{abstract}


%

\section{Introduction}
\input{Introduction.tex}
\section{Tensor Algebra and Notation}
\input{TensorAlgebraandNotation.tex}
\input{MIMO.tex}
\section{Simulation Results}
\input{SimulationResults.tex}
\section{Conclusion and Discussion}
In this paper, we have first presented a tensor model for MIMO-OFDM systems using the generalized tensor contraction operator between a channel tensor and a transmit signal tensor and the new properties of \cca{this} contraction operator. The \cca{proposed} model is a very general and flexible way of describing the received signal in MIMO-OFDM systems for all subcarriers jointly. We have also proposed a new representation of the channel tensor using a 4-way tensor with a special BTD structure. The resulting tensor model of the received signal \ca{enables} the design of the traditional ZF receiver and \ca{facilitates the design of} an iterative LS receiver based on projections and a recursive version of the iterative receiver ILSP denoted by RLSP. These algorithms based on projections have \ca{a} better performance than the ZF receiver if the number of transmitted frames is large enough. Moreover, the accuracy of the ILSP algorithm depends on the rank of the transmitted symbol matrices. Therefore, its performance in terms of the SER depends on the chosen modulation order and the modulation scheme. Hence, the system can be modified such that only specific code words are used. Moreover, the recursive algorithm can be modified such that it exploits the channel correlation in time varying scenarios. Note that we already exploit the correlation of the channel among adjacent subcarriers that leads to \ca{a} reduced number of pilot symbols as compared to other tensor models.

Next, we \ca{use} Khatri-Rao coding for the transmission of the OFDM symbols leading to Khatri-Rao coded MIMO-OFDM systems. The generalized tensor model using the contraction operator has been extended to the Khatri-Rao coded MIMO-OFDM system \ca{in a straightforward way}. In this case, the transmit signal tensor has a CP structure. By exploiting the overall tensor model, we propose a receiver based on \ca{the} LSKRF. This receiver requires the same amount of training symbols as traditional OFDM techniques, but it has an improved performance in terms of the SER. Hence, we benefit from the additional tensor structure of the transmitted signal \ca{to achieve a tensor gain}. In addition, we \ccb{have} propose\cca{d} to improve the performance of this receiver \ccb{further} by means of an additional LS iteration. Note that the Khatri-Rao coding strategy has \ca{a} reduced spectral efficiency than the uncoded MIMO-OFDM system. Therefore, we propose an alternative transmission technique where the "coding matrices" contain random data symbols. Thereby, this transmission technique also imposes a CP structure on the transmit signal tensor. Using the resulting received signal tensor, we \ccb{have} propose\cca{d} two receivers for randomly coded MIMO-OFDM systems. The first proposed receiver RC-KR estimates the symbol matrices based on \ca{the} LSKRF. The second proposed RC-KR+ALS receiver is an ALS algorithm initialized with the estimates of the symbol matrices using the RC-KR receiver. The proposed RC-KR+ALS algorithm outperforms the iterative receivers for MIMO-OFDM because it exploits the additional tensor structure of the signal tensor. Unlike the receivers for Khatri-Rao coded MIMO-OFDM, both receivers for the randomly coded MIMO-OFDM assume that ${M_{\rm{R}}\geq M_{\rm{T}}}$. However, the randomly coded system has \ca{a} higher spectral efficiency than the Khatri-Rao coded system.

\cca{{In the future}, \cca{for the Khatri-Rao coded MIMO-OFDM system} we can consider not just one additional LS iteration, but several iterations leading to an ALS based receiver initialized using the LSKRF. We should {also} consider the design of optimal orthogonal pilot sequences specific to the KR receiver. {Moreover, we can consider recursive LS instead of LS in order to relax this condition.} The randomly coded MIMO-OFDM system can be modified such that both symbol matrices contain symbols from different constellations and/or different modulation {orders}. This will lead to a resulting transmit signal tensor with diverse entries and potentially improved performance for the receivers in terms of {the} SER. {We should investigate which combinations of modulation orders and modulation types are suitable for  different SNRs.} {Furthermore, \ccb{the} general tensor model for multi-carrier systems proposed in this paper can be extended to other multi-carrier techniques such as UFMC and FBMC. Even more,  this model can be extended to relay-assisted systems and multi-user systems\ccb{, which would imply modeling the received signal tensor as a train of tensors linked by a sequence of double contractions, each hop representing one double contraction}.} \ccb{Another promising perspective is to assume} a low-rank structure of the channel tensor\ccb{, which will be a more realistic assumption for millimeter wave MIMO-OFDM  systems}. Exploiting \ccb{the double contraction} model and the additional structure of the low-rank \ccb{channel} tensor \ccb{would lead} to new blind receivers.}


%

\appendix
\input{ChannelTensor.tex}


\section*{Acknowledgment}

\ccb{This work has been supported by CAPES/PROBRAL Proc. no. 88887.144009/2017-00 and CAPES/PRINT Proc. no. 88887.311965/2018-00. }

\ifCLASSOPTIONcaptionsoff
  \newpage
\fi



\bibliographystyle{IEEEtran}
\bibliography{mybib}
\end{document}

%% file: macros.tex
\usepackage{bm,MnSymbol}

\renewcommand{\rm}{\textrm}

\newcommand{\ma}[1]{\bm{{#1}}}         
\newcommand{\ten}[1]{\ma{{\mathcal{#1}}}} 

\newcommand{\compl}{\mathbb{C}}        
\newcommand{\real}{\mathbb{R}}         

\newcommand{\opof}[2]{\mathop{{\rm{#1}}}\left(#2\right)}         

\newcommand{\diagof}[1]{\opof{diag}{#1}}         
\newcommand{\blockdiagof}[1]{\opof{blkdiag}{#1}} 

\newcommand{\vecof}[1]{\opof{vec}{#1}}           

\newcommand{\vecelem}[2]{#1_{(#2)}}    
\newcommand{\matelem}[3]{#1_{(#2,#3)}} 
\newcommand{\tenelem}[4]{#1_{(#2,#3,#4)}} 

\newcommand{\normof}[2]{\left\|#1\right\|_{#2}}
\newcommand{\honorm}[1]{\normof{#1}{\rm H}}               

\newcommand{\krp}{\diamond}  
\newcommand{\kron}{\otimes}  
\newcommand{\schur}{\odot} 	 
\newcommand{\ischur}{\oslash}  

\newcommand{\inv}{{-1}}          
\newcommand{\pinv}{+}          
\newcommand{\trans}{{\rm T}}   
\newcommand{\herm}{{\rm H}}    

\newcommand{\unfnot}[2]{\left[ #1 \right]_{(#2)}}    
\newcommand{\unf}[2]{\unfnot{\ten{#1}}{#2}}          

\newcommand{\cont}[2]{{\bullet}_{#1}^{#2}}   

\usepackage{color}

\newcommand{\cla}[1]{{\color{black} #1}}
\newcommand{\clb}[1]{{\color{black} #1}}
\newcommand{\clc}[1]{{\color{black} #1}}

\newcommand{\clg}[1]{{\color{black} #1}}
\newcommand{\clh}[1]{{\color{black} #1}}
\newcommand{\cli}[1]{{\color{black} #1}}
\newcommand{\ca}[1]{{\color{black} #1}}

\newcommand{\cca}[1]{{\color{black} #1}}
\newcommand{\ccb}[1]{{\color{black} #1}}
\newcommand{\ccc}[1]{{\color{black} #1}}
\newcommand{\ccd}[1]{{\color{black}#1}}
\newcommand{\cce}[1]{{\color{black} #1}}


%% file: Introduction.tex
In many tensor applications, we only have an element-wise or a slice-wise description of our
data/signal model. For instance, there exist only a slice-wise description of the PARATUCK2
decomposition and the PARAFAC2 decomposition corresponding to a certain unfolding of the overall tensor~\cite{HL96,H72}. In the same way, some proposed tensor based models for MIMO-ODFM communication systems have only an element-wise or a slice-wise representation~\cite{AFX13}. Further examples include the slice-wise description of MIMO communication systems \cca{using two-way relaying}~\cite{ZNNH15,XFdAS14}. This description of the signal models does not reveal the tensor structure explicitly. Hence, the derivation of all tensor unfoldings is not always obvious. Therefore, we propose to express the slice-wise multiplication of two tensors in terms of the \cca{double} contraction operator \cca{and use it do derive an explicit tensor structure of the received data tensor.}

OFDM is the most widely used multi-carrier technique in current wireless  communication systems. It is robust  in  multipath propagation  environments and has a simple and efficient implementation \cite{HYWL09}, \cite{B11}. Using the FFT (Fast Fourier Transform), the complete frequency band is divided into smaller frequency subcarriers. Moreover, the use of the  cyclic prefix mitigates  the ISI (Inter{-}Symbol Interference) and the ICI (Inter-Carrier Interference). Typically, the OFDM receiver is implemented in the frequency domain based on a ZF (Zero Forcing) filter. Other more advanced solutions are proposed in \cite{SFFM99}, as well as optimal training and channel estimation for OFDM systems are proposed in \cite{BLM03}, \cite{HYSH06}.

Tensor based signal processing offers an improved identifiability, uniqueness, and more efficient denoising compared to matrix based techniques. In \cite{AFX13}, a MIMO multi-carrier system is modeled using tensor algebra and the PARATUCK2 tensor decomposition resulting in a novel space, time, and frequency coding structure. Similarly in \cite{dAF12}, trilinear coding in space, time, and frequency is proposed for MIMO-OFDM systems based on the CP tensor decomposition. By exploiting tensor models, semi-blind receivers are introduced  for multi-carrier communications systems in \cite{FA14} and \cite{LCSA13}. All these \cca{works} use additional spreading that leads to a significantly reduced spectral efficiency to create the tensor structure. Moreover, previous publications on tensor models for multi-carrier communication systems \cite{AFX13}, \cite{dAF12}, \cite{LCSA13}, and \cite{FA14} do not exploit the channel correlation between the adjacent subcarriers. The previously mentioned publications {rely} on the subcarrier-wise description of the MIMO-OFDM system. This description of the signal models does not reveal the tensor structure explicitly. Hence, the derivation of all tensor unfoldings is not always obvious. Therefore, we propose to express the slice-wise multiplication of two tensors in terms of the \cca{double} contraction operator. \cca{To this end}, we summarize \cca{important} properties of the contraction operator for element-wise and slice-wise multiplications. Using the contraction operator, we \cca{derive the} tensor structure of the received signal that includes the frequency (subcarrier) mode.

In this paper, we first present the \cca{double} contraction between an uncoded signal tensor and a channel tensor for OFDM systems, yielding the same spectral efficiency as matrix based approaches (since no additional spreading is used) \cite{NHdA18}. By exploiting this new tensor structure, we can reshape the received signal tensor into the factorization of a sum of {Khatri}-Rao products. \cca{Channel and symbol estimation can be achieved} by means of \ccb{an} iterative and recursive least squares \ccb{procedure} originally proposed for blind source separation. {Moreover, we propose} an application of the \cca{double} contraction operator to Khatri-Rao coded MIMO-OFDM systems~\cite{NHdA17}.  Due to the Khatri-Rao coding, \cca{the signal tensor has more structure, i.e.,} we can use a CP model to describe \cca{it}. The Khatri-Rao space-time coding \cca{has been} introduced in \cite{SB02}. Later, it \cca{has been} extended in \cite{dAF13} to Khatri-Rao space-time-frequency coding. {In contrast to the state-of-the-art, we exploit} the structure of the channel {and the contraction properties using} the transmit signal tensor {and} the known coding matrix {to} propose a receiver based on the LS-KRF. In addition, we reduce the number of required pilot symbols by exploiting the correlation of the channel in the frequency domain. \cca{Alternatively}, we propose \ccb{a more spectrally efficient} "random coding" \ccb{model} for MIMO-OFDM systems. \cca{In this case,} we propose to keep the CP structure of the Khatri-Rao coded transmit signal~\cite{NHdA17} but the "coding matrix" contains \ccb{useful information} symbols. Thus, the proposed randomly coded MIMO-OFDM system has \cca{a} higher spectral efficiency than a Khatri-Rao coded MIMO-OFDM system. By exploiting the derived tensor structure of the received signal, we also \cca{design} two types of receivers for \cca{the} randomly coded MIMO-OFDM systems.

%% file: TensorAlgebraandNotation.tex
\subsection{Notation}
We use the following notation. Scalars are denoted either as capital or lower-case italic letters, $A, a$. Vectors and matrices, are denoted as bold-faced lower-case and capital letters, $\ma{a}, \ma{A}$, respectively.  Tensors are represented by bold-faced calligraphic letters $\ten{A}$. The following superscripts, $^\trans$, $^\herm$,$^{\inv}$, and $^\pinv$ denote transposition, Hermitian transposition, matrix inversion and Moore-Penrose pseudo matrix inversion, respectively. \ca{The outer product, Kronecker product, and Khatri-Rao product are denoted as $\circ$, $\otimes$, and $\diamond$, respectively.} \cca{Moreover, we denote an inverse Hadamard product (element-wise division) between two 	matrices of equal dimensions as $\ischur$.} The operators $\left|\left|.\right|\right|_{\text{F}}$ and $\left|\left|.\right|\right|_{\text{H}}$ denote the Frobenius norm and the higher order norm, respectively. Similar to matrices, a Kronecker product between two tensors $\ten{A} \in\compl^{M\times N \times L}$ and $\ten{A} \in\compl^{P\times Q \times R}$ can be defined as $\ten{K}=\ten{A}\kron\ten{B} \in\compl^{PM\times QN \times RL}$~\cite{C14}. Moreover, {the} $n$-mode product between a tensor $\ten{A} \in \compl^{I_1\times I_2 \ldots \times I_N}$ and a matrix  $\ma{B} \in \compl^{J \times I_n}$ is denoted as $\ten{A}\times_n\ma{B}$, for $n=1, 2, \ldots N $ \cite{KB09}. A super-diagonal or identity $N$-way tensor of dimension $R\times R\ldots \times R$ is denoted as $\ten{I}_{N,R}$. {Similarly}, an identity matrix {of} dimension ${R\times R}$ {is denoted} as $\ma{I}_R$ and {we denote} a vector of ones {of} length $R$ as $\ma{1}_R$. The $n$-th 3-mode slice of a tensor $\ten{A} \in \compl^{I\times J\times N}$ is denoted as $\ten{A}_{(.,.,n)}$ and accordingly one element of this tensor is denoted as $\ten{A}_{(i,j,n)}$. The operator ${\rm {diag}}(.)$ transforms a vector into a diagonal matrix and the operator ${\rm {vec}}(.)$ transforms a matrix into a vector.

\subsection{The CP Decomposition and Generalized Tensor Unfoldings}
The CP tensor decomposition decomposes a given tensor into the minimum number of rank one components. \cca{The} CP decomposition of a 4-way, rank $R$ noiseless tensor $\ten{A} \in \compl^{I \times J \times M \times N}$ {is defined as}
\begin{align} 
\ten{A} = \ten{I}_{3,R}\times_1\ma{F}_1\times_2\ma{F}_2\times_3\ma{F}_3\times_4\ma{F}_4, \label{cp4way} 
\end{align}
where $\ma{F}_1 \in \compl^{I \times R}, \ma{F}_2 \in \compl^{J \times R}$, $\ma{F}_3 \in \compl^{M \times R}$, and $\ma{F}_4 \in \compl^{N \times R}$ are the factor matrices \cite{KB09,CMdL15}. In addition to the $n$-mode unfoldings, generalized matrix unfoldings can be defined {by using} two subsets of any of the $N$ dimensions \cite{LA11,RSH12}. For instance, the set of modes $(1, 2,\ldots,N)$ of \clb{an} $N$-way tensor $\ten{A}$ can be divided into two \cla{non}-overlapping, $P$ and $N-P$ dimensional subsets, $\alpha^{(1)}=[\alpha_1 \ldots \alpha_P]$ and $\alpha^{(2)}=[\alpha_{P+1} \ldots \alpha_N]$, respectively. This leads to the generalized unfolding $\unfnot{\ten{A}}{\alpha^{(1)},\alpha^{(2)}}$, where the indices contained in $\alpha^{(1)}$ vary along the rows and the indices contained in $\alpha^{(2)}$ vary along the columns. Here, the index $\alpha_1$ varies the fastest between the rows, the index $\alpha_{P+1}$ varies the fastest between the columns, $P$ is any number between one and $N$, and $\alpha_{n}$ is any of the tensor dimensions. For instance, let us assume the 4-way tensor $\ten{A} \in \compl^{I \times J \times M \times N}$ defined in equation~\eqref{cp4way}. In the generalized unfolding $[\ten{A}]_{([1,2],[3,4])}$ the 1-st mode varies faster than the 2-nd mode along the rows and the 3-rd mode varies faster than the 4-th mode along the columns. Moreover, for a tensor with a CP structure, its unfoldings and generalized unfoldings can be expressed in terms of \cca{the} factor matrices. For instance, the generalized unfolding $[{\ten{A}}]_{([1, 2],[3, 4])}$ of the tensor $\ten{A}$ satisfies~\cite{RSH12,NHdA17}
\begin{align*}
[{\ten{A}}]_{([1, 2],[3, 4])} =\left(\ma{F}_2\diamond\ma{F}_1\right)\cdot\left(\ma{F}_4\diamond\ma{F}_3\right)^\trans.
\end{align*}
In {a} similar way{,} the rest of the tensor unfoldings and generalized unfoldings can be defined.

\subsection{Tensor Contraction}
The contraction $\ten{A}\bullet_n^m\ten{C}$ between two tensors  $\ten{A} \in \compl^{I_1\times I_2 \ldots \times I_N}$ and  $\ten{C} \in \compl^{J_1\times J_2 \ldots \times J_N}$  represents an inner product of the $n$-th mode of $\ten{A}$ with the  $m$-th mode of $\ten{C}$, provided that $I_n = J_m$ \cite{C14}. Contraction along several modes of compatible dimensions is also possible and accordingly the contraction along two modes is denoted as $\ten{A}\bullet_{n,k}^{m,l}\ten{C}$. \ccb{More specifically, the double contraction} between {the tensors} $\ten{A} \in \compl^{I \times J \times M \times N}$ and $\ten{C} \in \compl^{M \times N \times K}$ is defined as \cite{C14}{,}
	\begin{align*}
	(\ten{A}\bullet_{3,4}^{1,2}\ten{C})_{(i,j,k)} \triangleq \sum_{n=1}^{N}{\sum_{m=1}^{M}\ten{A}_{(i,j,m,n)}\cdot \ten{C}_{(m,n,k)}}=\ten{T}_{(i,j,k)}. 
	\end{align*}
 This example represents a contraction of the 3-rd and 4-th mode of $\ten{A}$ with the 1-st and 2-nd mode of $\ten{C}${,} respectively.

Using the concept of the generalized unfoldings, it can be shown that the tensor contraction satisfies
	\begin{align}
	[\ten{A}\bullet_{3,4}^{1,2}\ten{C}]_{([1,2],3)} &= [\ten{A}]_{([1,2],[3,4])}\cdot[\ten{C}]_{([1,2],3)} =	\label{ContractionTOGenUn1}
	\\ [\ten{A}\bullet_{4,3}^{2,1}\ten{C}]_{([1,2],3)} & = [\ten{A}]_{([1,2],[4,3])}\cdot[\ten{C}]_{([2,1],3)}. \label{ContractionTOGenUn2}
	\end{align} 
\ca{In} the generalized unfolding $[\ten{A}]_{([1,2],[3,4])}$ the 1-st mode varies faster than the 2-nd mode between the rows and the 3-rd mode varies faster then the 4-th mode between the columns. 

\subsection{Contraction Properties for Element-wise and Slice-wise Multiplications}
\subsubsection{{Hadamard product via tensor contraction}}
First, let us consider a Hadamard product (element-wise multiplication) between two vectors $\ma{a} \in \compl^{M \times 1}$ and  $\ma{b} \in \compl^{M \times 1}$, $\vecelem{c}{m}=\vecelem{a}{m}\vecelem{b}{m}$, $\forall m=1,\ldots, M$ ($\ma{c} \in \compl^{M \times 1}$). The Hadamard product can be expressed via the multiplication of a diagonal matrix and a vector, i.e., $\ma{a} \schur \ma{b} =  \diagof{\ma{a}}\ma{b} = \diagof{\ma{b}}\ma{a}$. Using \cca{the fact} that a matrix multiplication is equivalent to the contraction~$\cont{2}{1}$, we get
\begin{align*}
\ma{a} \schur \ma{b} =  \diagof{\ma{a}}\cont{2}{1}\ma{b} = \diagof{\ma{b}}\cont{2}{1}\ma{a}.
\end{align*}
%

\begin{figure}[htb]
	\centering
	\includegraphics[width=8cm]{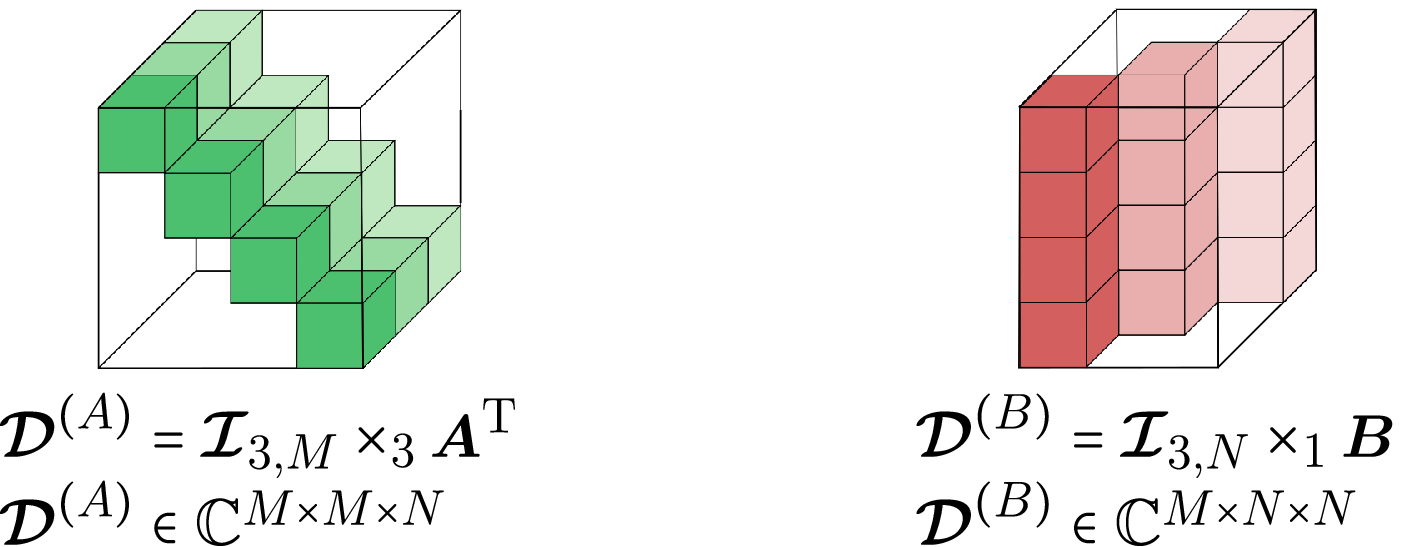}
	\caption{\cca{A visualization of the diagonal structure of the unfoldings for the tensors ${{\ten{D}^{(A)}} = \ten{I}_{3,M} \times_3 \ma{A}^\trans \in \compl^{M\times M\times N}}$ and ${\ten{D}^{(B)}} = \ten{I}_{3,N} \times_1 \ma{B}\in \compl^{M\times N\times N}$, for $M=4$ and $N=3$.}}
	\label{fig:BlockDiagonalD_AandD_B}
\end{figure}

Next, for the Hadamard product between two matrices $\ma{A} \in \compl^{M \times N}$ and  $\ma{B} \in \compl^{M \times N}$, $\matelem{\ma{C}}{m}{n}=\matelem{\ma{A}}{m}{n}\matelem{\ma{B}}{m}{n}$, $\forall m=1,\ldots, M$ and $n=1,\ldots, N$, we can show that $\ma{C} = \ma{A} \schur \ma{B} =  \ten{D}_A\cont{2,4}{1,2}\ma{B} = \ten{D}_B\cont{2,4}{1,2}\ma{A}$. Here $\ten{D}_A \in \compl^{M \times M \times N \times N}$ and $\ten{D}_B \in \compl^{M \times M \times N \times N}$ are diagonal {4-way} tensors with non-zero elements  $\tenelem{{\ten{D}_A}}{m,m}{n}{n} = \matelem{\ma{A}}{m}{n}$ and $\tenelem{{\ten{D}_B}}{m,m}{n}{n} = \matelem{\ma{B}}{m}{n}$, respectively. As an alternative, we also have    
\begin{align*}
\ma{C} = \ma{A} \schur \ma{B} =  \ten{D}^{(A)}\cont{2,3}{1,3}\ten{D}^{(B)}, 
\end{align*}
where the diagonal {3-way} tensors have the following non-zero elements $\tenelem{{\ten{D}^{(A)}}}{m}{m}{n}=\matelem{\ma{A}}{m}{n}$ and $\tenelem{{\ten{D}^{(B)}}}{m}{n}{n}=\matelem{\ma{B}}{m}{n}$. Moreover, these diagonal {3-way} tensors can be either defined it terms of slices, 
\begin{align*}
&\tenelem{{\ten{D}^{(A)}}}{.}{.}{n}=\diagof{{\matelem{\ma{A}}{.}{n}}}, \forall n = 1,\ldots, N 
\\&\tenelem{{\ten{D}^{(B)}}}{m}{.}{.}=\diagof{{\matelem{\ma{B}}{m}{.}}}, \forall m = 1,\ldots, M 
\end{align*}
or using tensor notation ${\ten{D}^{(A)}} = \ten{I}_{3,M} \times_3 \ma{A}^\trans$ and ${\ten{D}^{(B)}} = \ten{I}_{3,N} \times_1 \ma{B}$. \cca{The diagonal structure of these tensors is visualized in Fig.~\ref{fig:BlockDiagonalD_AandD_B}.} 

\subsubsection{{Slice-wise multiplication via tensor contraction}}
\begin{figure}[htb]
	\centering
	\includegraphics[width=1\linewidth]{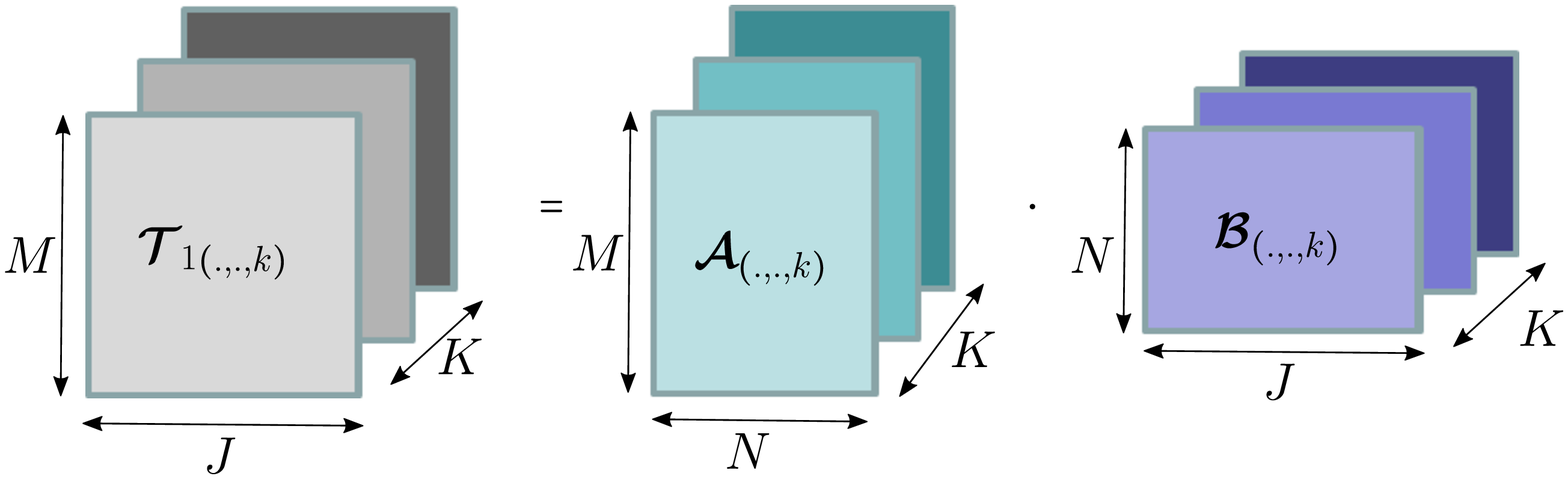}
	\caption{A slice-wise multiplication between two tensors $\ten{A} \in \compl^{M \times N \times K}$ and $\ten{B} \in \compl^{N \times J \times K}$.}
	\label{fig:HadamardProduct}
\end{figure}

A slice-wise multiplication between two tensors $\ten{A} \in \compl^{M \times N \times K}$ and $\ten{B} \in \compl^{N \times J \times K}$ is defined as $\tenelem{{\ten{T}_1}}{.}{.}{k} = \tenelem{\ten{A}}{.}{.}{k}\tenelem{\ten{B}}{.}{.}{k}$, $\forall k = 1,\ldots, K$.  We depict this slice-wise multiplication in Fig.~\ref{fig:HadamardProduct}. To express this slice-wise multiplication we can diagonalize $\ten{B}$ to obtain
\begin{align*}
\ten{T}_1 = \ten{A}\cont{2,3}{1,4}\ten{D}_B \in \compl^{M\times J \times K},
\end{align*}
where $\ten{D}_B\in \compl^{N\times J \times K \times K}$ has non-zero elements $\tenelem{{\ten{D}_B}}{n}{j}{k,k} = \tenelem{{\ten{B}}}{n}{j}{k}$ or $\tenelem{{\ten{D}_{B}}}{n}{j}{.,.}=\diagof{\tenelem{{\ten{B}}}{n}{j}{.}}$, for $n =1,\ldots N$ and $j =1,\ldots J$. Further combinations are also possible that lead to the same result, for instance{,} $\ten{T}_2 = \ten{D}_B\cont{1,4}{2,3}\ten{A} \in \compl^{J\times K \times M}$ or $\ten{T}_3 = \ten{D}_A\cont{2,4}{1,3}\ten{B} \in \compl^{M\times K \times J}$ with $\tenelem{{\ten{D}_A}}{m}{n}{k,k} = \tenelem{{\ten{A}}}{m}{n}{k}$ as diagonal elements (non-zero elements of $\ten{D}_A$). Note that the tensors $\ten{T}_1$, $\ten{T}_2$, and $\ten{T}_3$ {contain the same elements, but} have permuted dimensions. However, the permuted order of the dimensions is not relevant\clb{,} because  we always explicitly declare which dimension is multiplied or unfolded. 

\subsubsection{{Representation of diagonal matrices and diagonal tensors in terms of Khatri-Rao \clc{products}}}
An explicit expression of the diagonalized tensor can be obtained by expressing its generalized unfolding in terms of a {Khatri}-Rao product with an identity matrix. First, let us consider the column vector $\ma{a} \in\compl^{M}$. It can be easily shown that 
\begin{align*}
\diagof{\ma{a}} = \ma{I}_M\krp \ma{a}^\trans.
\end{align*}
\ccb{Next}, let us consider the reshaping of the matrix $\ma{A}\in\compl^{M\times N}$ into a diagonal tensor ${\ten{D}^{(A)}} = \ten{I}_{3,M} \times_3 \ma{A}^\trans$. By studying the resulting tensor structure, the tensor unfoldings, and the properties of the {Khatri}-Rao product, we get 
\begin{align*}
{\unfnot{\ten{D}^{(A)}}{[3,2],[1]}}= \ma{I}_M \diamond {{\ma{A}}}^\trans.
\end{align*}
Likewise, for the tensor ${\ten{D}^{(B)}} = \ten{I}_{3,N} \times_1 \ma{B}\in \compl^{M\times N\times N}$ and the matrix $\ma{B}\in \compl^{ M \times N}$, we have ${\unfnot{\ten{D}^{(B)}}{[1,3],[2]}}= \ma{I}_N \diamond {{\ma{B}}}$. 

\begin{table}[h]
	\begin{tabular}{l|l}  
		non-zero elements & generalized unfoldings \\
		\hline\hline
		 $\matelem{\ma{D}}{m}{m} = \vecelem{\ma{a}}{m}$ & $\ma{D} = \ma{I}_M \krp \ma{a}^\trans$  \rule{0pt}{2.6ex}  \\ 
		
		 $\tenelem{\ten{D}}{m}{n}{n} = \matelem{\ma{A}}{m}{n}$ & $\unf{\ten{D}}{[1,3],[2]} = \ma{I}_N \krp {\ma{A}}$  \rule{0pt}{2.6ex} \\ 
		
		 $\tenelem{\ten{D}}{m}{m}{n} = \matelem{\ma{A}}{m}{n}$ & $\unf{\ten{D}}{[3,2],[1]} = \ma{I}_M \krp {\ma{A}}^\trans$  \rule{0pt}{2.6ex} \\ 
		
		 $\tenelem{\ten{D}}{m,m}{n}{n} = \matelem{\ma{A}}{m}{n}$ & $\unf{\ten{D}}{[1,3],[2,4]} = \ma{I}_M \krp \vecof{\ma{A}}^\trans$  \rule{0pt}{2.6ex} \\ 
		
		$\tenelem{\ten{D}}{m,n}{k}{k} = \tenelem{\ma{A}}{m}{n}{k}$ & $\unf{\ten{D}}{[1,2,4],[3]} = \ma{I}_K \krp \unf{\ten{A}}{[1,2],[3]}$  \rule{0pt}{2.6ex} \\ 
		
	$\tenelem{\ten{D}}{m,m}{n}{k} = \tenelem{\ma{A}}{m}{n}{k}$ & $\unf{\ten{D}}{[3,4,2],[1]} = \ma{I}_M \krp \unf{\ten{A}}{[2,3],[1]}$  \rule{0pt}{2.6ex} \\ 
		
	\end{tabular}
	\centering
	\caption{{Link between the diagonalized tensor structures and their generalized unfoldings.}}
	\label{tqab}
\end{table}
The expression of the diagonalized tensor in terms of its generalized unfolding\ccb{s} and the Khatri-Rao product with an identity matrix can {also} be obtained for $N$-way tensors. Hence, there exists a link between the diagonalized tensor structures and their \ccb{corresponding} generalized unfoldings. {The \ccb{corresponding} generalized unfolding can always be expressed as a Khatri-Rao product between an identity matrix and a generalized unfolding of the {tensor to be diagonalized}, where the dimensions that are diagonalized are in the columns of the second matrix.} This notation will be used later in this paper and it is given in Table~\ref{tqab}.

The element-wise or slice-wise multiplication between two arrays (vectors/matrices/tensors) of the same order can be written in terms of a contraction if the {unaffected} mode vectors are transformed into a diagonal matrix (by adding an additional array dimension). {This} diagonalization can be performed using the Khatri-Rao product as shown in Table~\ref{tqab}.

%% file: MIMO.tex
\section{MIMO-OFDM} \label{OFDM}
We assume a MIMO-OFDM system with $M_{\rm T}$ transmit and $M_{\rm R}$ receive antennas. One OFDM block consists of $N$ samples, which equals the DFT (Discrete Fourier Transform) length, using the assumption that all $N$ subcarriers are used for data transmission. If guard subcarriers are used, i.e., not all subcarries are used for data transmission, the number of OFDM samples is smaller that the DFT length. All signals and equations used for the following derivation are in the frequency domain. Moreover, $N$ is the number of subcarriers and $K$ denotes the number of transmitted frames. The received signal in the frequency domain $\tilde{\ten{Y}} \in\compl^{N\times M_{\rm R} \times K}$ after the removal of the cyclic prefix is defined by means of the contraction operator
\begin{align}
\tilde{\ten{Y}} = \tilde{\ten{H}}\bullet_{2,4}^{1,2}\tilde{\ten{S}} + \tilde{\ten{N}} = \tilde{\ten{Y}}_0 + \tilde{\ten{N}}. \label{RxSigContraction}
\end{align}
We use $\sim$ to distinguish the frequency domain from the time domain, i.e., $\tilde{\ten{Y}}={\ten{Y}}\times_1\ma{F}_N$, where $\ma{F}_N\in \compl^{N\times N}$ is the DFT matrix and $\ten{Y}$ is the received signal in \cca{the} time domain. The transmit signal tensor is denoted as $\tilde{\ten{S}}  \in\compl^{N \times M_{\rm T} \times K}$ and $\tilde{\ten{N}}  \in\compl^{N \times M_{\rm R} \times K}$ represents the additive white Gaussian noise in the frequency domain. The tensor $\tilde{\ten{Y}}_0 \in\compl^{N\times M_{\rm R} \times K}$ represents the noiseless received signal in \ccb{the} frequency domain after the removal of the cyclic prefix. The frequency-selective propagation channel is represented by a channel tensor  $\tilde{\ten{H}}  \in\compl^{N\times N \times M_{\rm R} \times M_{\rm T}}$ as we propose in \cite{NHdA17} \ccb{the structure of which is detailed} as follows.

\subsection{Channel tensor}
\begin{figure}[!t]
	\centering
	\includegraphics[width=0.6\linewidth]{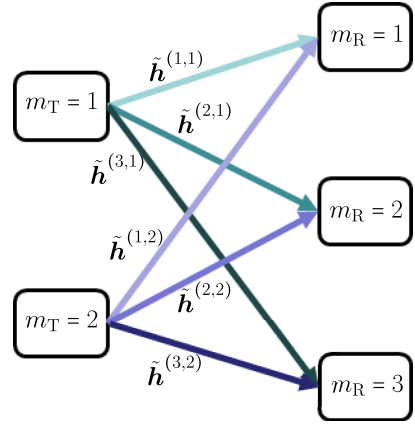}
	\caption{A MIMO system with $M_{\rm T}=2$ transmit antennas and $M_{\rm R}=3$ receive antennas.}
	\label{fig:MIMOSystem}
\end{figure}

\ccb{We assume that the} frequency-selective channel has an impulse response $\ma{h}_L^{(m_{\rm R},m_{\rm T})} \in \compl^{L \times 1}$, for each receive-transmit antenna pair, $(m_{\rm R},m_{\rm T})$, \ccb{for $m_{\rm R} = 1\ldots M_{\rm R}$ and  $m_{\rm T} = 1\ldots M_{\rm T}$,} and a maximum of $L$ taps. After the removal of the cyclic prefix, the channel matrix in the frequency domain is a diagonal matrix for each receive-transmit antenna pair, $\tilde{\ma{H}}^{(m_{\rm R},m_{\rm T})}={\rm{ diag}}\left(\ma{F}_{N\times L}\cdot \ma{h}_L^{(m_{\rm R},m_{\rm T})}\right) \in \compl^{N \times N}$ \cite{BLM03,HYSH06}. Here, the matrix $\ma{F}_{N\times L} \in \compl^{N \times L}$ contains the first $L$ columns of the DFT matrix of size $N \times N$. Collecting all the channel matrices in a 4-way channel tensor $\tilde{\ten{H}}$, we get
\begin{align}
&\tilde{\ten{H}}_{(.,.,m_{\rm R},m_{\rm T})} = {\rm{ diag}}\left(\ma{F}_{N\times L}\cdot \ma{h}_L^{(m_{\rm R},m_{\rm T})}\right)={\rm{ diag}}\left(\tilde{\ma{h}}^{(m_{\rm R},m_{\rm T})}\right) \label{ChannelTensor}. 
\end{align}
For each receive-transmit antenna pair the channel transfer matrix is a diagonal matrix that is represented by the corresponding slice of the tensor $\tilde{\ten{H}}$ as shown in \eqref{ChannelTensor}. The vector $\tilde{\ma{h}}^{(m_{\rm R},m_{\rm T})} \in \compl^{N\times1}$ contains the frequency domain channel coefficients. An example of a MIMO system with $M_{\rm T}=2$ transmit antennas and $M_{\rm R}=3$ receive antennas and the corresponding channel vectors \cca{is} depicted in Fig.~\ref{fig:MIMOSystem}. We assume that the channel stays constant during the $K$ frames. Note that only in case of cyclic prefix OFDM the channel tensor in the frequency domain contains diagonal matrices for each receive-transmit antenna pair. In a general multi-carrier system, the frequency domain channel matrix is not necessarily diagonal. However, equation \eqref{RxSigContraction} is still satisfied which means \ccb{that our general model also remains} valid for non-orthogonal multi-carrier systems.

In $\eqref{ChannelTensor}$, we have defined the channel tensor{. However, \cca{up to this point,}} we have {not} revealed the explicit tensor structure. In order to do so, let us first assume that all channel transfer matrices for the $m_{\rm T}$-th transmit and all receive antennas are collected in a diagonal tensor $\tilde{\ten{H}}_{\rm R}^{(m_{\rm T})} \in \compl^{N\times N \times M_{\rm R}}$, \cca{i.e.,}
\begin{align}
&\tilde{\ten{H}}_{{\rm R}(.,.,m_{\rm R})}^{(m_{\rm T})} = {\rm {diag}}\left(\tilde{\ma{h}}^{(m_{\rm R},m_{\rm T})}\right) \label{eq:ChannelTensorPerTX} \\&m_{\rm R} = 1,\ldots, M_{\rm R}, m_{\rm T} = 1,\ldots, M_{\rm T} \notag
\end{align}
Based on this diagonal structure, the tensor $\tilde{\ten{H}}_{\rm R}^{(m_{\rm T})}$ \ccb{can be written as} the following CP decomposition
\begin{align}
\tilde{\ten{H}}_{\rm R}^{(m_{\rm T})} = \ten{I}_{3,N}\times_1\ma{I}_N\times_2\ma{I}_N\times_3\tilde{\ma{H}}_{\rm R}^{(m_{\rm T})}, \label{eq:ChannelTensorPerTXCP}
\end{align}
where
$\tilde{\ma{H}}_{\rm R}^{(m_{\rm T})} =\begin{bmatrix}
\tilde{\ma{h}}^{(1,m_{\rm T})} &  \tilde{\ma{h}}^{(2,m_{\rm T})} & \ldots & \tilde{\ma{h}}^{(M_{\rm R},m_{\rm T})}
\end{bmatrix}^\trans \in \compl^{M_{\rm R} \times N}$.

\begin{figure}[!t]
	\centering
	\includegraphics[width=3cm]{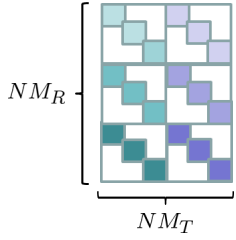}
	\caption{\ca{Visualization of the generalized unfolding $[\tilde{\ten{H}}]_{([1,3],[2,4])}$ for a MIMO-OFDM system with parameters $M_{\rm T} = 2$, $M_{\rm R} = 3$, and $N = 3$.}}
	\label{fig:BlockDiagonalChannelStructure}
\end{figure}

The complete 4-way channel tensor, defined in equation \eqref{ChannelTensor} can be obtained by concatenating the $\tilde{\ten{H}}_{\rm R}^{(m_{\rm T})}$ tensors along the fourth dimension. Hence, the 4-way channel tensor $\tilde{\ten{H}}$ can be expressed as
\begin{align}
\tilde{\ten{H} } &=\begin{bmatrix}
\tilde{\ten{H}}_{\rm R}^{(1)} & \sqcup_4 & \tilde{\ten{H}}_{\rm R}^{(2)} &  \sqcup_4 & \ldots & \tilde{\ten{H}}_{\rm R}^{(M_{\rm T})}
\end{bmatrix} \notag
\\ & = \sum_{m_{\rm T}=1}^{M_{\rm T}} \tilde{\ten{H}}_{\rm R}^{(m_{\rm T})} \circ \ma{e}_{m_{\rm T}} \\ &= \sum_{m_{\rm T}=1}^{M_{\rm T}} \ten{D}\times_1\ma{I}_N\times_2\ma{I}_N\times_3\tilde{\ma{H}}_{\rm R}^{(m_{\rm T})}\times_4 \ma{e}_{m_{\rm T}}. \label{eq:ChannelTensorModel}
\end{align}
Note that $\tilde{\ten{H}}$ satisfies a very special BTD, where $\ten{D}_{(.,.,.,1)}=\ten{I}_{3,N} \in \real^{N\times N\times N\times 1}$ ($\ten{D}=\ten{I}_{4,1}\kron\ten{I}_{3,N}$) and $\ma{e}_{m_{\rm T}}\in \real^{M_{\rm T}\times 1}$ is a pining vector. We prove the BTD structure of the channel tensor $\tilde{\ten{H}}$ in Appendix~\ref{App:Channel Tensor}. In \cca{this appendix}, we also show that the ${([1,3],[2,4])}$ generalized unfolding of the channel tensor can be expressed as
\begin{align}
[\tilde{\ten{H}}]_{([1,3],[2,4])} = \tilde{\ma{H}}\diamond(\ma{1}_{M_{\rm T}}^\trans\otimes\ma{I}_{N}) \in \compl^{NM_{\rm R} \times NM_{\rm T}}, \label{eq:ChannelTensorUnfolding}
\end{align}
where $\tilde{\ma{H}} \in \compl^{ M_{\rm R} \times NM_{\rm T}}$ is a matrix containing all non-zero elements of the tensor $\tilde{\ten{H}}$ and it is defined as,
\begin{align}
\tilde{\ma{H}} &=
\left[
\begin{array}{c|c|c|c}
\tilde{\ma{h}}^{(1,1)\trans} & \tilde{\ma{h}}^{(1,2)\trans} & \ldots & \tilde{\ma{h}}^{(1,M_{\rm T})\trans}\\
\vdots & \vdots &\vdots & \vdots \\
\tilde{\ma{h}}^{(M_{\rm R},1)\trans} & \tilde{\ma{h}}^{(M_{\rm R},2)\trans} & \ldots & \tilde{\ma{h}}^{(M_{\rm R},M_{\rm T})\trans}
\end{array}
\right] \notag
\\&=
\left[
\begin{array}{cccc}
\tilde{\ma{H}}^{(1)}_{\rm R} & \tilde{\ma{H}}^{(2)}_{\rm R} & \ldots & \tilde{\ma{H}}^{(M_{\rm T})}_{\rm R}
\end{array}
\right] \in \compl^{ M_{\rm R} \times NM_{\rm T}}. \label{Hstructure}
\end{align}
\ca{Fig.~\ref{fig:BlockDiagonalChannelStructure} depicts the structure of the generalized unfolding $[\tilde{\ten{H}}]_{([1,3],[2,4])}$ for a MIMO-OFDM system with parameters $M_{\rm T} = 2$, $M_{\rm R} = 3$, and $N = 3$.}

\subsection{Data transmission}
The signal tensor $\tilde{\ten{S}}$ in equation~\eqref{RxSigContraction} contains all data symbols in the frequency domain that are transmitted on $N$ subcarriers, $M_{\rm T}$ transmit antennas, and $K$ frames. For notational simplicity, we define the following block matrix $\tilde{\ma{S}}$ as the transpose of the 3-mode unfolding of $\tilde{\ten{S}}$
\begin{align}
\tilde{\ma{S}} = \tilde{\ten{S}}_{([1,2],[3])}^\trans =
\left[
\begin{array}{cccc}
\tilde{\ma{S}}^{(1)} & \tilde{\ma{S}}^{(2)} & \ldots & \tilde{\ma{S}}^{(M_{\rm T})}
\end{array}
\right] \in \compl^{K \times NM_{\rm T}}, \label{Sstructure}
\end{align}
where $\tilde{\ma{S}}^{(m_{\rm T})} \in \compl^{K \times N}$ contains the symbols transmitted via the $m_{\rm T}$-th antenna.

Moreover, we assume that the symbol matrix consists of data and pilot symbols, $\tilde{\ma{S}}= \tilde{\ma{S}}_{\rm{d}}+\tilde{\ma{S}}_{\rm{p}}$. The {matrices} $\tilde{\ma{S}}_{\rm{d}}$ and $\tilde{\ma{S}}_{\rm{p}}$ represent the data symbols and the pilot symbols, respectively. The matrix $\tilde{\ma{S}}_{\rm{d}}$ contains zeros at the positions of the pilot symbols. Accordingly, the matrix $\tilde{\ma{S}}_{\rm{p}}$ contains non-zero elements only at the pilot positions. Typically, there are three ways of arranging the pilot symbol within the OFDM blocks (block, comb, and lattice-type) \cite{HYWL09}. We assume \clh{a} comb-type arrangement of the pilot symbols with equidistant positions in the time and the frequency domain\cca{s}, for each antenna. The spacing in the time domain is denoted by $\Delta K$. Moreover, we \cca{assume a} subcarrier spacing of $\Delta F$ between two pilot symbols. Furthermore, there are positions where neither pilot symbols \cca{nor} data symbols are allowed to be transmitted. These positions are reserved for the pilot symbols corresponding to the remaining antennas. This results in $M_{\rm T}\lfloor\frac{N}{\Delta F}\rfloor$ pilot symbols per frame. In comparison, other publications such as \cite{AFX13}, \cite{dAF12}, \cite{LCSA13}, and \cite{FA14} use $N\clh{M_{\rm T}}$ pilot symbols per frame. {By exploiting the channel correlation among adjacent subcarriers,  a reduced number of pilot symbols can be used for channel estimation.}

\subsection{Receiver design}
Using the property of the generalized unfoldings in equation~\eqref{ContractionTOGenUn1}, the received signal in equation~\eqref{RxSigContraction} becomes
\begin{align}
[\tilde{\ten{Y}}]_{([1, 2],[3])} = [\tilde{\ten{H}}]_{([1,3],[2,4])}\tilde{\ten{S}}_{([1,2],[3])} + [\tilde{\ten{N}}]_{([1, 2],[3])}. 
\label{RXsignal}
\end{align}
Next, by substituting the corresponding tensor unfoldings in equation \eqref{RXsignal}, we get
\begin{align}
[\tilde{\ten{Y}}]_{([1, 2],[3])} = \left(\tilde{\ma{H}}\diamond(\ma{1}_{M_{\rm T}}^\trans\otimes\ma{I}_{N})\right)\cdot\tilde{\ma{S}}^\trans + [\tilde{\ten{N}}]_{([1, 2],[3])}. \label{RXUnfoding}
\end{align}
The above equation satisfies an unfolding of a noisy observation of a low-rank tensor with a CP structure. By applying an inverse unfolding for the received signal in the frequency domain after the removal of the cyclic prefix, we get \ccb{the desired tensor description of the received data tensor}
\begin{align}
\tilde{\ten{Y}} = \ten{I}_{3,NM_{\rm T}} \times_1 (\ma{1}_{M_{\rm T}}^\trans\otimes\ma{I}_{N}) \times_2 \tilde{\ma{H}} \times_3 \tilde{\ma{S}} + \tilde{\ten{N}} \ca{\in\compl^{N\times M_{\rm R} \times K}}. \label{RXCP}
\end{align}
Our goal is to jointly estimate the channel and the symbols, i.e., $\tilde{\ma{H}}$ and $\tilde{\ma{S}}$ in equation \eqref{RXCP}. Note that all factor matrices are flat\cca{,} resulting in a degenerate CP model in all three modes. Therefore, it is difficult to estimate the channel and the symbols by simply fitting a CP model \cca{to} the received signal tensor in \eqref{RXCP}.

Using the prior knowledge of the pilot symbols and their positions, the channel in the frequency domain can be estimated. Naturally, the channel is estimated only at those subcarrier positions where the pilot symbols are located. Afterwards, an interpolation is applied to get the complete channel estimate. \ccb{Alternatively}, as shown in \cite{BLM03,HYSH06} the channel can be first estimated in the time domain and then transformed into the frequency domain. Either way, this leads to a pilot based channel estimate that we denote as ${\tilde{\ma{H}}}_{\rm{p}}$, or $\tilde{\ten{H}}_{\rm{p}}$\footnote{\clh{In our simulations, we use the pilot based channel estimate obtained in the time domain.}}. The pilot based channel estimate is then used to estimate the data symbols.  In the remainder of this section, we discuss different ways to estimate the symbols. We use the pilot based channel estimate to initialize the proposed algorithms.

Traditionally, the estimate of the symbols is obtained in the frequency domain with a ZF receiver. In this case, the symbols are calculated by inverting the channel matrix for each subcarrier individually. This ZF receiver using the above defined tensor notation is summarized in Algorithm~\ref{ZF}. 
{\tiny
\begin{algorithm}[h!]

	\SetAlgoLined
	\DontPrintSemicolon
	initialization   $\tilde{\ten{H}}_{\rm{p}}$\;
	\For{$n = 1:N$}{
		$\tilde{{\ten{S}}}_{(n,.,.)}\approx {\tilde{\ten{H}}}_{{\rm{p}}(n,n,.,.)}^+{\tilde{\ten{Y}}}_{(n,.,.)}$\;	
	}
	\KwResult{{$\tilde{{{\ten{{S}}}}}$}}
	\caption{ZF receiver} \label{ZF}

\end{algorithm}
}

Alternatively, if we compute the 1-mode unfolding of the tensor ${\tilde{\ten{Y}}}$ in equation \eqref{RXCP}, we get
\begin{align*}
[\tilde{\ten{Y}}]_{([1],[2, 3])} = (\ma{1}_{M_{\rm T}}^\trans\otimes\ma{I}_{N})\cdot\left(\tilde{\ma{S}}\diamond\tilde{\ma{H}}\right)^\trans +[\tilde{\ten{N}}]_{([1],[3, 2])}. 
\end{align*}
Taking into account the structure of the matrices $(\ma{1}_{M_{\rm T}}^\trans\otimes\ma{I}_{N}) \in \real^{N \times NM_{\rm T}}$, $\tilde{\ma{H}}$ in \eqref{Hstructure}, and $\tilde{\ma{S}}$ in \eqref{Sstructure}, \cca{the 1-mode} unfolding becomes
\begin{align*}
[\tilde{\ten{Y}}]_{([1],[2, 3])} = \sum_{m_{\rm T}=1}^{M_{\rm T}}\left(\tilde{\ma{S}}^{(m_{\rm T})}\diamond\tilde{\ma{H}}_{\rm R}^{(m_{\rm T})}\right)^\trans +[\tilde{\ten{N}}]_{([1],[3, 2])}.
\end{align*}
After transposition and omitting the noise term, we get
\begin{align*}
[\tilde{\ten{Y}}]_{([2, 3],[1])} \approx \sum_{m_{\rm T}=1}^{M_{\rm T}}\left(\tilde{\ma{S}}^{(m_{\rm T})}\diamond\tilde{\ma{H}}_{\rm R}^{(m_{\rm T})}\right) \in \compl^{M_{\rm R}K  \times N}.
\end{align*}
This sum of Khatri-Rao products can be resolved in a column-wise fashion. Let $\tilde{\ma{y}}_n\in \compl^{M_{\rm R}K \times 1}$ denote the $n$-th column of $[\tilde{\ten{Y}}]_{([2, 3],[1])} \in \compl^{M_{\rm R}K \times N}$. After reshaping this vector into \ccb{the} matrix $\tilde{\ma{Y}}_n \in \compl^{M_{\rm R}\times K}$, such that $\tilde{\ma{y}}_n={\rm {vec}}(\tilde{\ma{Y}}_n)$, it is easy to see that this matrix satisfies
\begin{align}
\tilde{\ma{Y}}_n \approx \tilde{\ma{H}}_n\cdot\tilde{\ma{S}}_n, \label{Problem}
\end{align}
where $\tilde{\ma{H}}_n$ and $\tilde{\ma{S}}_n$ are the $n$-th slices of $\tilde{\ten{H}}_{(n,n,.,.)}\in \compl^{M_{\rm R} \times M_{\rm T}}$ and $\tilde{\ten{S}}_{(n,.,.)}\in \compl^{M_{\rm T} \times K}$, respectively. Note that $\tilde{\ma{Y}}_n$ is the $n$-th slice of ${\tilde{\ten{Y}}}_{(n,.,.)}$. Using the pseudo inverse of the channel, we get the traditional ZF receiver as summarized in Algorithm~\ref{ZF}.

\ccb{Alternatively}, the channel and the symbols on each subcarrier can be estimated by means of iterative or recursive LS algorithms. Similar algorithms were proposed in  \cite{TVP94} and \cite{TVP96} for blind source separation \clh{on \cli{a} single subcarrier}. We extend two of the algorithms presented in \cite{TVP96} that are based on projection to our application. We \ccb{have proposed} an extension of these algorithm using enumeration in~\cite{NHdA18}. In this paper, \cca{our focus is} on the algorithms using projection \ccb{since} that they are computationally less expensive than the algorithms based on enumeration.

\begin{algorithm}[h!]
	\SetAlgoLined
	\DontPrintSemicolon
	initialization $\tilde{\ten{H}}_{\rm{p}}$, maxIteration, minErr\;
	\For{$n = 1 : N$}{
		set $i= 1$, $e=\infty$\;
		\While{$i<$~maxIteration or $e <$ minErr}{
			$\bar{\tilde{\ma{S}}}_n^{(i)}= ({\tilde{\ma{H}}}_n^{(i-1)\herm} {\tilde{\ma{H}}}_n^{(i-1)})^{-1}{\tilde{\ma{H}}}_n^{(i-1)\herm}\tilde{\ma{Y}}_n$\;
			${\tilde{\ma{S}}}_n^{(i)}={\rm {proj}}\left(\bar{\tilde{\ma{S}}}_n^{(i)}\right)$\;
			\eIf{${\rm{rank}}\left({\tilde{\ma{S}}}_n^{(i)}\right)=M_{\rm T}$}{
				${\tilde{\ma{H}}}_n^{(i)}= \tilde{\ma{Y}}_n{\tilde{\ma{S}}}_n^{(i)\herm}({\tilde{\ma{S}}}_n^{(i)} {\tilde{\ma{S}}}_n^{(i)\clh{\herm}})^{-1}$\;
			}{
				${\tilde{\ma{H}}}_n^{(i)}={\tilde{\ma{H}}}_n^{(i-1)}$\;
			}
			$i = i+1$,
			$e = \| {\tilde{\ma{H}}}_n^{(i-1)}- {\tilde{\ma{H}}}_n^{(i)}\|_{\rm F}^2$\;
		}
	}
	\KwResult{${\tilde{\ten{S}}}$ and $\tilde{\ten{H}}$}
	\caption{Iterative Least-Squares with Projection (ILSP)} \label{ILSP}
\end{algorithm}

The algorithm ILSP (Iterative Least-Squares with Projection) summarized in Algorithm~\ref{ILSP} is an iterative \cca{solution} based on LS. It is initialized with the pilot based channel estimate, the maximum number of iterations (maxIteration), and the minimum error difference between two consecutive updates (minErr). The ILSP algorithm is essentially an iterative version of the ZF algorithm, where in each iteration the estimated symbols are projected onto the finite alphabet $\Omega$ of the transmitted symbols. This finite alphabet depends on the modulation type and the modulation order $M_o$. Details regarding the convergence for different finite alphabets are discussed in \cite{TVP96}. To estimate the symbols, we compute a pseudo inverse of the channel which leads to the condition $M_{\rm R}\geq M_{\rm T}$. The algorithm \ccb{conditionally} updates the channel \ccb{depending on the rank of the estimated symbol matrix, i.e., the channel is updated} if the rank of the symbol matrix $\tilde{\ma{S}}_n \in \compl^{M_{\rm T}\times K}$ is $M_{\rm T}$, {$K\geq M_{\rm T}$}. Note that this is not possible for all values of $M_{\rm T}$, $K$, and for all patterns of random data symbols {from a finite distribution}.

\begin{algorithm}[h]
	\SetAlgoLined
	\DontPrintSemicolon
	initialization $\tilde{\ten{H}}_{\rm{p}}$, $0\leq \alpha \leq 1$\;
	\For{$n = 1 : N$}{
		$\bar{\tilde{\ma{S}}}_n= ({\tilde{\ma{H}}}_n^{\herm} {\tilde{\ma{H}}}_n)^{-1}{\tilde{\ma{H}}}_n^{\herm}\tilde{\ma{Y}}_n$\;
		${\tilde{\ma{S}}}_n={\rm {proj}}\left(\bar{\tilde{\ma{S}}}_n \right)$\;
		set $\ma{P}^{(0)} = \ma{I}_{M_{\rm T}}$, ${\tilde{\ma{H}}}_n^{(0)}=\tilde{\ma{H}}_n$, $\alpha=1$\;
		\For{$k = 1:K$}
		{	${\ma{s}} = {\tilde{\ma{S}}}_{n(.,k)}$\;
			${\tilde{\ma{H}}}_n^{(k)}= {\tilde{\ma{H}}}_n^{(k-1)}+ \frac{\left(\tilde{\ma{Y}}_{n(.,k)}-{\tilde{\ma{H}}}_n^{(k-1)}{\ma{s}}\right)}{\alpha+ {\ma{s}}^\herm\ma{P}^{(k-1)}{\ma{s}}}{\ma{s}}^\herm\ma{P}^{(k-1)}$\;\;
			$\ma{P}^{\prime(k)}= \frac{1}{\alpha}\left( \ma{P}^{\prime (k-1)}-\frac{\ma{P}^{\prime(k-1)}\ma{s}\ma{s^\herm}\ma{P}^{\prime(k-1)}}{\alpha+\ma{s}^\herm\ma{P}^{\prime(k-1)}\ma{s}}\right)$ \;
		}
	}
	\KwResult{${\tilde{\ten{S}}}$ and $\tilde{\ten{H}}$}
	\caption{Recursive Least-Squares with Projection (RLSP)} \label{RLSP}
\end{algorithm}

The second algorithm, namely RLSP (Recursive Least-Squares with Projections) is a recursive implementation of ILSP. {We summarize this algorithm in Algorithm~\ref{RLSP}.} The channel is estimated based on RLS (Recursive Least Squares), where $\alpha$ is the weighting coefficient and $\ma{P}^\prime$ denotes the inverse correlation matrix. Due to the computation of the pseudo-inverse of the channel matrix for the algorithm RLSP, $M_{\rm R}\geq M_{\rm T}$ should hold.

ILSP has the same computational complexity as traditional ZF receivers with the added complexity of the additional iterations if the symbol matrix has full rank.  The RLSP \cca{algorithm} requires a finite number of iterations that is equal to $NK$. 
\cca{Taking into account that \ccb{the} multiplication of two \ccb{complex} matrices \ccb{of} sizes $m \times n$ and $n \times l$ requires approximately $2nml$ operations and an inversion of a matrix of size $r\times r$ requires $\frac{2}{3}r^3$ operations, the ZF algorithm requires  $\frac{2}{3}M_\rm{T}^3 + 4M_\rm{T}^2M_\rm{R} + 2M_\rm{T}M_\rm{R}^2K$ operations per subcarrier. The ILSP algorithm requires $\frac{4}{3}M_\rm{T}^3 + 6M_\rm{T}^2M_\rm{R} + 2M_\rm{T}M_\rm{R}^2K + 2M_\rm{T}^2K + 2M_\rm{T}M_\rm{R}K$ operations per subcarrier and iteration. Finally, the RLSP algorithm requires $\frac{2}{3}M_\rm{T}^3 + 4M_\rm{T}^2M_\rm{R} + 2M_\rm{T}M_\rm{R}^2K + 2M_\rm{R}M_\rm{T}K + 2M_\rm{T}^2M_\rm{R}K+ 2M_\rm{T}^2K$ operations per subcarrier and iteration.} \ccb{Therefore, the ILSP algorithm is less complex then the RLSP if the number of antennas is small. However, with the increase of the number of antennas the RLSP algorithm is less complex than the ILSP algorithm.}

\section{Khatri-Rao coded MIMO-OFDM} \label{KR-OFDM}
In this section, we model a Khatri-Rao coded MIMO-OFDM communication system as a \cca{double} tensor contraction between a channel and a signal tensor that contains coded symbols. This \cca{double} tensor contraction is essentially equivalent to the model in \eqref{RxSigContraction}. However, we assume that the signal tensor contains Khatri-Rao coded symbols.

As in the previously presented MIMO-OFDM model without coding (see Section~\ref{OFDM}), we assume a MIMO-OFDM communication system with $M_{\rm T}$ transmit and $M_{\rm R}$ receive antennas. One OFDM block consists of $N$ samples, which equals the DFT length. Moreover, all $N$ subcarriers are used for data transmission. Furthermore, we assume a frequency-selective channel model that stays constant over the transmission of $P$ frames. In contrast to the model presented in Section~\ref{OFDM}, here, we assume that the $P$ frames are divided into $K$ groups of $Q$ blocks \ca{($Q$ corresponds to the spreading factor)}, $P = K\cdot Q$.

Accordingly, the received signal in the frequency domain is given by
\begin{align}
\tilde{\ten{Y}} = \tilde{\ten{H}}\bullet_{2,4}^{1,2}\tilde{\ten{X}} + \tilde{\ten{N}} =  \tilde{\ten{Y}}_0+ \tilde{\ten{N}} \in \compl^{N \times M_{\rm R}\times K\times Q}, \label{RxSigContraction2}
\end{align}
where $\tilde{\ten{H}} \in \compl^{N \times N \times M_{\rm R} \times M_{\rm T}}$ is the channel tensor and $\tilde{\ten{X}} \in \compl^{N \times M_{\rm T} \times K \times Q}$ is the signal tensor. The tensor $\tilde{\ten{N}}\in \compl^{N \times M_{\rm R}\times K\times Q}$ contains additive white Gaussian noise and $\tilde{\ten{Y}}_0\in \compl^{N \times M_{\rm R}\times K\times Q}$ is the noiseless received signal.

\subsection{Channel tensor}
In this section, we use the model of the channel tensor $\tilde{\ten{H}}$ defined in equation~\eqref{eq:ChannelTensorModel}. Moreover, we have defined the generalized unfolding $\unfnot{\tilde{\ten{H}}}{[1,3],[2,4]}$ in equation~\eqref{eq:ChannelTensorUnfolding}. Using a permutation matrix, it can be shown that the generalized unfolding $[\tilde{\ten{H}}]_{([1,3],[4,2])}$ of the channel is equal to
\begin{align}
[\tilde{\ten{H}}]_{([1,3],[4,2])} = \bar{\ma{H}} \diamond (\ma{I}_N\otimes\ma{1}_{M_{\rm T}}^\trans), \label{GUChannel}
\end{align}
where \begin{align*}
\bar{\ma{H}} = \underbrace{\begin{bmatrix}
	\tilde{\ma{H}}_R^{(1)} & \ldots & \tilde{\ma{H}}_{\rm R}^{(M_{\rm T})}
	\end{bmatrix}}_{\tilde{\ma{H}}}\cdot\ma{P} \in \compl^{M_{\rm R} \times M_{\rm T}N}.
\end{align*}
The permutation matrix $\ma{P} \in \real^{NM_{\rm T} \times M_{\rm T}N }$ reorders the columns such that the faster increasing index is $M_{\rm T}$ instead of $N$ and it is defined as $[\tilde{\ten{H}}]_{([1,3],[4,2])}=[\tilde{\ten{H}}]_{([1,3],[2,4])}\cdot\ma{P}$. \cca{Recall} that the matrices $\tilde{\ma{H}}\in \compl^{M_{\rm R} \times NM_{\rm T}}$ and $\tilde{\ma{H}}_{\rm R}^{(m_{\rm T})}\in \compl^{M_{\rm R} \times N}$ are defined in equation~\eqref{Hstructure}. The structure of the 4-way channel tensor in the frequency domain $\tilde{\ten{H}}$ and its unfoldings are derived in Appendix~\ref{App:Channel Tensor}.

\subsection{Data transmission}
We can impose a CP structure \cca{to} the transmit signal tensor, if we assume Khatri-Rao coded symbols \cite{SB02,dAF13}. The coding is proportional to the number of transmit antennas if we use a spreading factor $Q=M_{\rm T}$, for each subcarrier $n = 1,2,\ldots, N$. Hence, the generalized unfolding of the signal tensor is
\begin{align}
[\tilde{\ten{X}}]_{([2,1],[4,3])} &= \begin{bmatrix}
\tilde{\ma{S}}_1\diamond\ma{C}_1 & \tilde{\ma{S}}_2\diamond\ma{C}_2 & \ldots &\tilde{\ma{S}}_N\diamond\ma{C}_N
\end{bmatrix}^\trans \notag \\&= \ma{I}_{ M_{\rm T}N}(\bar{\ma{S}}\diamond\bar{\ma{C}})^\trans, \label{KRSymbols}
\end{align}
where the matrix $\tilde{\ma{S}}_n \in \compl^{K \times M_{\rm T}}$ contains modulated data symbols and  $\ma{C}_n \in \compl^{Q \times M_{\rm T}}$ is a Vandermonde coding matrix as defined in \cite{SB02}. The matrices $\bar{\ma{S}} = \begin{bmatrix}
\tilde{\ma{S}}_1 & \ldots & \tilde{\ma{S}}_N
\end{bmatrix} \in \compl^{K\times M_{\rm T}N}$ and
$\bar{\ma{C}} = \begin{bmatrix}
\ma{C}_1 & \ldots & \ma{C}_N
\end{bmatrix} \in \compl^{Q\times M_{\rm T}N} $ contain all symbol and coding matrices for each subcarrier, respectively. Note that $\bar{\ma{S}}=\tilde{\ma{S}}\cdot\ma{P}$, where the matrix $\tilde{\ma{S}}$ is defined in equation~\eqref{Sstructure} and $\ma{P} \in \real^{NM_{\rm T} \times M_{\rm T}N }$ is the above mentioned permutation matrix that reorders the columns such that the faster increasing index is $M_{\rm T}$ instead of $N$. Moreover, we assume that $\tilde{\ma{S}}$ contains pilot symbols as explained after equation~\eqref{Sstructure}. As shown in \cite{SB02} and as directly follows from \eqref{KRSymbols}, the tensor $[\tilde{\ten{X}}]_{([2,1],3,4)}$ satisfies the following CP decomposition
\begin{align*}
[\tilde{\ten{X}}]_{([2,1],3,4)} = \ten{I}_{3,M_{\rm T}N} \times_1 \ma{I}_{M_{\rm T}N} \times_2 \bar{\ma{S}} \times_3 \bar{\ma{C}}.
\end{align*}

\subsection{Receiver Design}
Using equations \eqref{ContractionTOGenUn1}, \eqref{ContractionTOGenUn2}, and \eqref{RxSigContraction2} the noiseless received signal can be expressed as
\begin{align*}
[\tilde{\ten{Y}}_0]_{([1,2],[4,3])}=[\tilde{\ten{H}}]_{([1,3],[4,2])}\cdot[\tilde{\ten{X}}]_{([2,1],[4,3])}.
\end{align*}
Inserting the corresponding unfoldings of the channel and the signal tensor in equation \eqref{GUChannel} and \eqref{KRSymbols}, respectively, the noiseless received signal in the frequency domain is given by
\begin{align*}
[\tilde{\ten{Y}}_0]_{([1,2],[4,3])}= \left(\bar{\ma{H}} \diamond (\ma{I}_N\otimes\ma{1}_{M_{\rm T}}^\trans)\right)\cdot(\bar{\ma{S}}\diamond\bar{\ma{C}})^\trans.
\end{align*}
The above equation represents an unfolding of a 4-way tensor with a CP structure. Therefore, the noiseless received signal tensor can be expressed as
\begin{align}
\tilde{\ten{Y}}_0 = {\ten{I}_{4,M_{\rm T}N}}\times_1 (\ma{I}_N\otimes\ma{1}_{M_{\rm T}}^\trans) \times_2 \bar{\ma{H}} \times_3  \bar{\ma{S}}\times_4  \bar{\ma{C}} \ca{ \in \compl^{N \times M_{\rm R}\times K\times Q}}. \label{RXSignal}
\end{align}
Equation \eqref{RXSignal} represents the received signal in the frequency domain for all $N$ subcarriers, $M_{\rm R}$ receive antennas, and $P$ frames after the removal of the cyclic prefix. Depending on the available \cca{\textit{a priori}} knowledge at the receiver side, channel estimation, symbol estimation, or joint channel and symbol estimation can be performed.

Let us compare the MIMO-OFDM tensor model and the Khatri-Rao coded MIMO-OFDM tensor model in equations~\eqref{RXCP} and~\eqref{RXSignal}, respectively. First, the factor matrices in these equations have different index ordering\ccb{s}. In equation~\eqref{RXCP} the faster increasing index in $N$, whereas in equation~\eqref{RXSignal} the faster increasing index in $M_{\rm T}$ along the columns of the factor matrices. We use $\sim$ and $-$ to distinguish the different index orderings of the factor matrices. Recall that we have defined a permutation matrix $\ma{P}$ that considers the reordering of the columns of the factor matrices. Moreover, equation~\eqref{RXSignal} has an additional tensor dimension (the 4-mode) corresponding to the coding technique and the spreading factor $Q$. Furthermore, taking into account the permutation matrix $\ma{P}$, we get equation~\eqref{RXCP} from equation~\eqref{RXSignal} for $Q = 1$ and $\bar{\ma{C}}= \ma{1}_{M_{\rm T}N}^\trans$ (i.e., no coding and the spreading factor equals one).

Using equation \eqref{RXSignal}, the channel and the data symbols can be jointly estimated from the $([1,4],[3,2])$ generalized unfolding of the noise corrupted received signal
\begin{align*}
[\tilde{\ten{Y}}]_{([1,4],[3,2])}\approx\left( \bar{\ma{C}} \diamond (\ma{I}_N\otimes\ma{1}_{M_{\rm T}}^\trans)\right)\cdot(\bar{\ma{H}}\diamond\bar{\ma{S}})^\trans.
\end{align*}
Under the assumption that $Q=M_{\rm T}$, $\left( \bar{\ma{C}} \diamond (\ma{I}_N\otimes\ma{1}_{M_{\rm T}}^\trans)\right) \in \compl^{NQ \times M_{\rm T}N}$ is a block diagonal, left invertible matrix and known at the receiver. Using the properties of the coding matrices defined in \cite{SB02}, i.e., $\ma{C}_n^\herm\ma{C}_n=M_{\rm T}\ma{I}_{M_{\rm T}}$, we have
\begin{align*}
\bar{\ma{Y}}\triangleq\frac{1}{M_{\rm T}}\left(\bar{\ma{C}} \diamond (\ma{I}_N\otimes\ma{1}_{M_{\rm T}}^\trans)\right)^\herm\cdot[\tilde{\ten{Y}}]_{([1,4],[3,2])} \approx (\bar{\ma{H}}\diamond\bar{\ma{S}})^\trans.
\end{align*}
After transposition,
$\bar{\ma{Y}}^\trans \approx \bar{\ma{H}}\diamond\bar{\ma{S}}$ can be approximated by the Khatri-Rao product between the channel and the data symbols. Therefore, the channel and the data symbols can be jointly estimated based on the LSKRF as in~\cite{RH09b}.

Using the LSKRF, the matrices $\bar{\ma{H}}$ and $\bar{\ma{S}}$ can be identified up to one complex scaling factor ambiguity per column. Hence, the estimated matrices \cca{satisfy the following relations}
\begin{align}
&\hat{\bar{\ma{H}}} = \bar{\ma{H}}\cdot\ma{\Lambda},  \label{ScaledChannel}
\\ &\hat{\bar{\ma{S}}} = \bar{\ma{S}}\cdot\ma{\Lambda}^{-1}, \label{Symbols}
\end{align}
where $\ma{\Lambda} \in \compl^{M_{\rm T} N\times M_{\rm T}N}$ is a diagonal matrix with diagonal elements equal to the $M_{\rm T}N$ complex scaling ambiguities. The simplest way to resolve the scaling ambiguity is \cca{by assuming the knowledge of} one row of the matrix $\bar{\ma{S}} \in \compl^{K\times M_{\rm T}N}$. This corresponds to $M_{\rm T}N$ pilot symbols, \cca{i.e.,} one pilot symbol per transmit antenna and subcarrier. Since traditional MIMO-OFDM communication systems use less pilot symbols than $M_{\rm T}N$, we propose to use the same amount of pilot symbols and exploit the channel correlation between adjacent subcarriers in order to estimate the scaling matrix. We transmit pilot symbols on positions with equidistant spacing in the frequency and the time domain. 
With the \ca{prior} knowledge of the pilot symbols and their positions, we can obtain an initial channel estimate as in traditional MIMO-OFDM systems (see Section~\ref{OFDM}). We denote this pilot based channel estimate by  $\tilde{\ten{H}}_{\rm{p}}$ $({\bar{\ma{H}}}_p)$. The pilot based channel estimate is then used to estimate the scaling ambiguity $\ma{\Lambda}$ in equation~\eqref{ScaledChannel} as
\begin{align*}
\hat{\ma{\Lambda}} = {\rm {diag}}\left(\frac{1}{M_{\rm R}}\sum_{m_{\rm R}= 1}^{M_{\rm R}}\hat{\bar{\ma{H}}}_{(m_{\rm R},.)}\oslash{\bar{\ma{H}}}_{\rm{p}(m_{\rm R},.)}\right).
\end{align*}

By multiplying the solution of the LSKRF with the diagonal matrix $\hat{\ma{\Lambda}}$, the scaling ambiguity in equation \eqref{Symbols} is resolved and the data symbols can be demodulated.  Note that the proposed Khatri-Rao receiver estimates the channel and the symbols in a semi-blind fashion. First, the channel and the symbols are jointly estimated without any \cca{\textit{a priori}} information. The pilot based channel estimate is then used to resolve the scaling ambiguity affecting the columns of $\hat{\bar{\ma{H}}}$ and $\hat{\bar{\ma{S}}}$. Therefore, the optimal length and repetition of the piloting sequences are identical as for the traditional OFDM systems. We summarize the steps of the proposed Khatri-Rao (KR) receiver in Algorithm~\ref{Alg:KR}.
\begin{algorithm}[h]
	\SetAlgoLined
	\DontPrintSemicolon
	Initialization: $\tilde{\ten{H}}_{\rm{p}}$ and $\bar{\ma{C}}$;\;
	1. Compute $\bar{\ma{Y}}=\frac{1}{M_{\rm T}}\left(\bar{\ma{C}} \diamond (\ma{I}_N\otimes\ma{1}_{M_{\rm T}}^\trans)\right)^\herm\cdot[\tilde{\ten{Y}}]_{([1,4],[3,2])}$.\;\;
	2. Compute the LSKRF of $\bar{\ma{Y}}^\trans$ using the algorithm proposed in~\cite{RH09b} which gives $\hat{\bar{\ma{H}}}$ and $\hat{\bar{\ma{S}}}$;\;
	3. Compute the scaling matrix $\hat{\ma{\Lambda}} = {\rm {diag}}\left(\frac{1}{M_{\rm R}}\sum_{m_{\rm R}= 1}^{M_{\rm R}}\hat{\bar{\ma{H}}}_{(m_{\rm R},.)}\oslash{\bar{\ma{H}}}_{\rm{p}(m_{\rm R},.)}\right)$. The matrix ${\bar{\ma{H}}}_{\rm{p}(m_{\rm R},.)}$ is defined as in equation~\eqref{GUChannel} using the estimated channel tensor $\tilde{\ten{H}}_{\rm{p}}$;\; 
	4. Resolve the scaling ambiguity $\bar{\ma{H}}= \hat{\bar{\ma{H}}}\cdot\hat{\ma{\Lambda}}^\inv$ and $\bar{\ma{S}}= \hat{\bar{\ma{S}}}\cdot\hat{\ma{\Lambda}}$.\;
	\KwResult{${\bar{\ma{S}}}$ and $\bar{\ma{H}}$}
	\caption{Khatri-Rao (KR) receiver} \label{Alg:KR}
\end{algorithm}

Furthermore, the channel estimate resulting from the KR receiver can be used for channel tracking in future transmission frames if the channel has not changed drastically. If the channel estimate is used for tracking, it could be improved by means of an additional LS estimate from $[\tilde{\ten{Y}}]_{([2,4,1],[3])}$ with the knowledge of the estimated and projected symbols onto the finite alphabet $\Omega$, i.e., $\ca{Q({{\bar{\ma{S}}}})}={\rm {proj}}({{\bar{\ma{S}}}})$. The finite alphabet $\Omega$ depends on the modulation type and the modulation order $M_o$.
\begin{align*}
\hat{\bar{\ma{H}}}_{\rm {LS}} ^\trans=  \left((\ma{I}_N\otimes\ma{1}_{M_{\rm T}}^\trans)\diamond\bar{\ma{C}}\diamond \ca{Q({\bar{\ma{S}}})}\right)^{+}\cdot[\tilde{\ten{Y}}]_{([2,4,1],[3])}
\end{align*}
However, we can also use this improved channel estimation to \cca{further} improve the performance of the KR receiver. Using this updated channel estimate an improved estimate of the diagonal scaling matrix $\hat{\ma{\Lambda}}$ can be calculated and with that an enhanced estimate of the symbols, $\hat{\bar{\ma{S}}}_{\rm {LS}}$, using equation \eqref{Symbols}. Note that, instead of just one LS estimate of the channel and the symbols the performance can be \cca{further} enhanced with additional iterations leading to an iterative receiver. Note that the symbol matrix $\hat{\bar{\ma{S}}}_{\rm {LS}}$ can be estimated in the least squares sense from the 3-mode unfolding of equation \eqref{RXSignal}, but the estimation of $\hat{\ma{\Lambda}}$ is computationally cheaper. The KR receiver with its enhancement via LS is summarized in Algorithm~\ref{Alg:KR+LS}.
\begin{algorithm}[h]
	\SetAlgoLined
	\DontPrintSemicolon
	1. Apply Algorithm~\ref{Alg:KR}\;
	2. Project the symbols onto the finite alphabet $\Omega$, i.e.,  $\ca{Q({\bar{\ma{S}}})}={\rm {proj}}({\bar{\ma{S}}})$;\;
	3. Compute an enhanced channel estimate $\hat{\bar{\ma{H}}}_{\rm {LS}}^\trans=  \left((\ma{I}_N\otimes\ma{1}_{M_{\rm T}}^\trans)\diamond\bar{\ma{C}}\diamond \ca{Q({\bar{\ma{S}}})}\right)^{+}\cdot[\tilde{\ten{Y}}]_{([2,4,1],[3])}$;\;
	4. Improve the estimate of the diagonal scaling matrix $\hat{\ma{\Lambda}}_{\rm {LS}} = {\rm {diag}}\left(\frac{1}{M_{\rm R}}\sum_{m_{\rm R}= 1}^{M_{\rm R}}\hat{\bar{\ma{H}}}_{(m_{\rm R},.)}\oslash\hat{\bar{\ma{H}}}_{{\rm {LS}}(m_{\rm R},.)}\right)$;\;
	5. Compute an enhanced estimate of the symbol matrix $\hat{\bar{\ma{S}}}_{\rm {LS}} = {\bar{\ma{S}}}\cdot\hat{\ma{\Lambda}}_{\rm {LS}}$.
	\caption{Khatri-Rao receiver and its enhancement via Least-Squares (KR+LS)} \label{Alg:KR+LS}
	\KwResult{$\hat{\bar{\ma{S}}}_{\rm {LS}}$ and $\hat{\bar{\ma{H}}}_{\rm {LS}}$}
\end{algorithm}

Due to the additional LS based estimates the KR+LS algorithm has higher computational complexity than the KR algorithm.
\section{Randomly coded MIMO-OFDM} \label{RCKR-OFDM}
In Section~\ref{KR-OFDM}, we have proposed a tensor model for KR coded MIMO-OFDM systems that introduces an additional CP structure to the signal tensor.  The additional CP structure of the signal tensor is achieved by means of Khatri-Rao coding. However, using the Khatri-Rao coding, we add additional spreading that reduces the spectral efficiency of the system. Therefore, in this section we propose to keep the CP structure of the signal tensor proposed in Section~\ref{KR-OFDM}, but to introduce random coding. We introduce the random coding, \cca{which means that the} "coding matrix" \cca{also} contains data symbols.

As in Section~\ref{KR-OFDM}, the received signal in the frequency domain  after the removal of the cyclic prefix is given by
\begin{align}
\tilde{\ten{Y}} = \tilde{\ten{H}}\bullet_{4,2}^{2,1}\tilde{\ten{X}} + \tilde{\ten{N}} =  \tilde{\ten{Y}}_0+ \tilde{\ten{N}} \in \compl^{N \times M_{\rm R}\times K\times Q}, \label{RxSigContraction22}
\end{align}
where $\tilde{\ten{H}} \in \compl^{N \times N \times M_{\rm R} \times M_{\rm T}}$ is the channel tensor and $\tilde{\ten{X}} \in \compl^{N \times M_{\rm T} \times K \times Q}$ is the signal tensor. The tensor $\tilde{\ten{N}}$ contains additive white Gaussian noise and $\tilde{\ten{Y}}_0$ is the noiseless received signal. As for the KR coded MIMO-OFDM system, we transmit $P = KQ$ frames that are divided into $K$ groups of $Q$ blocks \ca{("spreading factor")}. The number of subcarriers is $N$, and $M_{\rm R}$ and $M_{\rm T}$ denote the number of receive and transmit antennas, respectively.
\subsection{Channel tensor}
We model the channel tensor $\tilde{\ten{H}}$ according to equation~\eqref{eq:ChannelTensorModel}. Details regarding this model are also provided in Appendix~\ref{App:Channel Tensor}. In this section\cca{,} we use the generalized unfolding $[\tilde{\ten{H}}]_{([1,3],[4,2])} = \bar{\ma{H}} \diamond (\ma{I}_N\otimes\ma{1}_{M_{\rm T}}^\trans)$ that is defined in~\eqref{GUChannel}.
\subsection{Data Transmission}
As previously mentioned, we impose a CP structure on the signal tensor $\tilde{\ten{X}}$ similar to the Khatri-Rao coding proposed in Section~\ref{KR-OFDM}. For the generalized unfolding $([2,1],[4,3])$ of the signal tensor, we have
\begin{align}
[\tilde{\ten{X}}]_{([2,1],[4,3])} &= \begin{bmatrix}
\ca{\bar{\ma{S}}_1}\diamond\ma{C}^\prime_1 & \ca{\bar{\ma{S}}_2}\diamond\ma{C}^\prime_2 & \ldots &\ca{\bar{\ma{S}}_N}\diamond\ma{C}^\prime_N
\end{bmatrix}^\trans \notag \\&= \ma{I}_{ M_{\rm T}N}(\bar{\ma{S}}\diamond\bar{\ma{C}}^\prime)^\trans, \label{KRSymbols2}
\end{align}
where the matrix $\ca{\bar{\ma{S}}_n} \in \compl^{K \times M_{\rm T}}$ contains modulated data symbols. \ca{In contrast to the Khatri-Rao coding in Section~\ref{KR-OFDM},} here, we assume that the first row of the matrix $\ma{C}_n^\prime \in \compl^{Q \times M_{\rm T}}$ contains only ones, whereas the remaining $Q-1$ rows contain modulated data symbols. Hence, the ``coding matrix" (the matrix $\ma{C}_n$ in~\eqref{KRSymbols} represents the coding matrix) contains also random entries. We refer to this transmission technique as random coding. Moreover, the matrices $\bar{\ma{S}} = \begin{bmatrix}
\ca{\bar{\ma{S}}_1} & \ldots & \ca{\bar{\ma{S}}_N}
\end{bmatrix} \in \compl^{K\times M_{\rm T}N}$ and
$\bar{\ma{C}}^\prime = \begin{bmatrix}
\ma{C}_1^\prime & \ldots & \ma{C}_N^\prime
\end{bmatrix} \in \compl^{Q\times M_{\rm T}N} $ contain symbol and random coding matrices for each subcarrier, respectively. Note that $\bar{\ma{S}}$ is defined as in Section~\ref{KR-OFDM}, i.e.,  $\bar{\ma{S}}=\tilde{\ma{S}}\cdot\ma{P}$, where the matrix $\tilde{\ma{S}}$ is defined in equation~\eqref{Sstructure} and $\ma{P} \in \real^{NM_{\rm T} \times M_{\rm T}N }$ is the permutation matrix that reorders the columns such that the faster increasing index is $M_{\rm T}$ instead of $N$. Moreover, we assume that $\tilde{\ma{S}}$ contains pilot symbols as explained after equation~\eqref{Sstructure}. As shown in \cite{SB02} and as directly follows from \eqref{KRSymbols2}, the tensor $[\tilde{\ten{X}}]_{([2,1],3,4)}$ satisfies the following CP decomposition 
\begin{align*}
[\tilde{\ten{X}}]_{([2,1],3,4)} = \ten{I}_{3,M_{\rm T} N} \times_1 \ma{I}_{M_{\rm T}N} \times_2 \bar{\ma{S}} \times_3 \bar{\ma{C}}^\prime.
\end{align*}
\subsection{Receiver Design}
Using equations \eqref{ContractionTOGenUn1} and \eqref{RxSigContraction22}, for the noiseless received signal, we get
\begin{align}
[\tilde{\ten{Y}}_0]_{([1,2],[4,3])}=[\tilde{\ten{H}}]_{([1,3],[4,2])}\cdot[\tilde{\ten{X}}]_{([2,1],[4,3])}. \label{eq:RXSig}
\end{align}
Inserting the corresponding unfoldings of the channel tensor and the signal tensor, i.e., inserting ~\eqref{GUChannel} and~\eqref{KRSymbols2} into~\eqref{eq:RXSig}, we \ccb{obtain}
\begin{align*}
[\tilde{\ten{Y}}_0]_{([1,2],[4,3])}= \left(\bar{\ma{H}} \diamond (\ma{I}_N\otimes\ma{1}_{M_{\rm T}}^\trans)\right)\cdot(\bar{\ma{S}}\diamond\bar{\ma{C}}^\prime)^\trans.
\end{align*}
The above equation represents an unfolding of a 4-way tensor with CP structure. Therefore, it can be expressed as
\begin{align}
\tilde{\ten{Y}}_0 = {\ten{I}_{4,M_{\rm T}N}}\times_1 (\ma{I}_N\otimes\ma{1}_{M_{\rm T}}^\trans) \times_2 \bar{\ma{H}} \times_3  \bar{\ma{S}}\times_4  \bar{\ma{C}}^\prime \ca{\in \compl^{N \times M_{\rm R}\times K\times Q}}. \label{RXSignal2}
\end{align}
Equation~\eqref{RXSignal2} represents the noiseless received signal in the frequency domain for all $N$ subcarriers, $M_{\rm R}$ receive antennas, and $P$ frames after the removal of the cyclic prefix for MIMO-OFDM system with RC (Random Coding). Note that the CP decomposition in~\eqref{RXSignal2} is degenerate in all four modes.

Depending on the available \cca{\textit{a priori}} knowledge at the receiver side, channel estimation, symbol estimation, or joint channel and symbol estimation can be performed. For instance, from the 3-mode unfolding of \ca{the} tensor $\ten{Y}_0$ in ~\eqref{RXSignal2}, we can obtain
\begin{align}
\bar{\ma{S}} = \unfnot{{\ten{Y}_0}}{3}\cdot\left[{\left(\bar{\ma{C}}^\prime\krp \bar{\ma{H}}\krp (\ma{I}_N\otimes\ma{1}_{M_{\rm T}}^\trans)\right)^\trans}\right]^\ca{\pinv}, \label{US}
\end{align}
provided that $M_{\rm R}Q\geq M_{\rm T}$.
Moreover, from the 4-mode unfolding and 2-mode unfolding of tensor $\ten{Y}_0$ in~\eqref{RXSignal2}, we can obtain $\bar{\ma{C}}^\prime$ and $\bar{\ma{H}}$, respectively.
\begin{align}
\bar{\ma{C}}^\prime = \unfnot{{\ten{Y}_0}}{4}\cdot\left[\left(\bar{\ma{S}}\krp \bar{\ma{H}}\krp (\ma{I}_N\otimes\ma{1}_{M_{\rm T}}^\trans)\right)^\trans\right]^\ca{\pinv} \label{UC}
\\ \bar{\ma{H}} = \unfnot{{\ten{Y}_0}}{2}\cdot\left[\left(\bar{\ma{C}}^\prime\krp \bar{\ma{S}}\krp (\ma{I}_N\otimes\ma{1}_{M_{\rm T}}^\trans)\right)^\trans\right]^\ca{\pinv} \label{UH}
\end{align}
Note that we can compute $\bar{\ma{C}}^\ca{\prime}$ \ca{via a pseudo-inverse} if $M_{\rm R}K\geq M_{\rm T}$. \cca{This condition is a design requirement for the proposed MIMO-OFDM system with random coding.}

For noisy observations such as~\eqref{RxSigContraction22}, the equations~\eqref{US}-\eqref{UH} hold approximately. In this case, we can use the equations~\eqref{US}-\eqref{UH} to estimates the symbols and the channel in an ALS fashion. However, there is no guarantee of convergence if we initialize the ALS algorithm randomly. Therefore, we propose to use the pilot based channel estimate $\bar{\ma{H}}_{\rm{p}}$ to obtain initial estimates of the matrices $\bar{\ma{S}}$ and $\bar{\ma{C}}^\prime$ based on LSKRF. This pilot based channel estimated is obtained from the pilot symbols in $\bar{\ma{S}}$ and the first row of $\bar{\ma{C}}^\prime$ that has entries equal to one. From the $([3,4],[1,2])$ generalized unfolding of the noisy observation $\ten{Y}$, we get
\begin{align*}
\unfnot{\tilde{\ten{Y}}}{[3,4],[1,2]} \approx \left[\bar{\ma{C}}^\prime\krp\bar{\ma{S}}\right]\cdot\left[\bar{\ma{H}}_{\rm{p}}\krp(\ma{I}_N\otimes\ma{1}_{M_{\rm T}}^\trans)\right]^\trans.
\end{align*}

\begin{algorithm}[h]
	\SetAlgoLined
	\DontPrintSemicolon
	Initialization: $\bar{\ten{H}}_{\rm{p}}$;\;
	1. Compute $\bar{\ma{Y}}=\unfnot{\ten{Y}}{[3,4],[1,2]}\cdot\left[\left(\bar{\ma{H}}_{\rm{p}}\krp(\ma{I}_N\otimes\ma{1}_{M_{\rm T}}^\trans)\right)^\trans\right]^\pinv$;\;
	2. Compute the LSKRF of $\bar{\ma{Y}}$ using the algorithm proposed in~\cite{RH09b} which gives $\hat{\bar{\ma{C}}}^\prime$ and $\hat{\bar{\ma{S}}}$;\;
	3. Compute the scaling matrix $\hat{\ma{\Lambda}} = {\rm {diag}}\left(\hat{\bar{\ma{C}}}^\prime_{(1,.)}\oslash{\bar{\ma{C}}}^\prime_{(1,.)}\right)$. (The first row of the matrix ${\bar{\ma{C}}}^\prime$ contains only ones);\;
	4. Resolve the scaling ambiguity $\bar{\ma{C}}^\prime= \hat{\bar{\ma{C}}}^\prime\cdot\hat{\ma{\Lambda}}^\inv$ and $\bar{\ma{S}}= \hat{\bar{\ma{S}}}\cdot\hat{\ma{\Lambda}}$;\;
	\KwResult{$\bar{\ma{S}}$ and $\bar{\ma{C}}^\prime$}
	\caption{Random Coding-Khatri-Rao (RC-KR) receiver} \label{Alg:RC-KR}
\end{algorithm}
\begin{algorithm}[h!]
	\SetAlgoLined
	\DontPrintSemicolon
	Apply Algorithm~\ref{Alg:RC-KR};\;
	\While{\clg{does not} exceed the maximum number of iterations, \clg{does not reach} a predifined minimum, or the error of the cost \clg{function} has not changed within two consecutive iteration\clg{s}}{
		\eIf{${\rm {rank}}\left(\left[\bar{\ma{C}}^\prime\krp \bar{\ma{S}}\krp (\ma{I}_N\otimes\ma{1}_{M_{\rm T}}^\trans)\right]^\trans\right)=M_{\rm T}N$}{ Update $\hat{\bar{\ma{H}}} = \unfnot{{\ten{Y}_0}}{2}\cdot\left[\left(\hat{\bar{\ma{C}}}^\prime\krp \hat{\bar{\ma{S}}}\krp (\ma{I}_N\otimes\ma{1}_{M_{\rm T}}^\trans)\right)^\trans\right]^\ca{\pinv}$\;
		}{keep the \ca{previous} estimate of $\hat{\bar{\ma{H}}}$;\;
		}\;
		{Update $\hat{\bar{\ma{C}}}^\prime = \unfnot{{\ten{Y}_0}}{4}\cdot\left[\left(\hat{\bar{\ma{S}}}\krp \hat{\bar{\ma{H}}}\krp (\ma{I}_N\otimes\ma{1}_{M_{\rm T}}^\trans)\right)^\trans\right]^\ca{\pinv}$;\;
			Project $\hat{\bar{\ma{C}}}^\prime={\rm{proj}}\left(\hat{\bar{\ma{C}}}^\prime\right)$ onto the finate alphabet $\Omega$;	
		}\;
		{Update $\hat{\bar{\ma{S}}} = \unfnot{{\ten{Y}_0}}{3}\cdot\left[\left(\hat{\bar{\ma{C}}}^\prime\krp \hat{\bar{\ma{H}}}\krp (\ma{I}_N\otimes\ma{1}_{M_{\rm T}}^\trans)\right)^\trans\right]^\ca{\pinv}$;\;
			Project $\ca{Q}(\hat{\bar{\ma{S}}})={\rm{proj}}\left(\hat{\bar{\ma{S}}}\right)$ onto the finate alphabet $\Omega$.
		}\; 		
	}
	\KwResult{$\hat{\bar{\ma{S}}}$, $\hat{\bar{\ma{C}}}^\prime$, and $\hat{\bar{\ma{H}}}$}
	\caption{Random Coding-Khatri-Rao + ALS (RC-KR+ALS) receiver} \label{Alg:RC-KR+ALS}
\end{algorithm}
Given $\bar{\ma{H}}_{\rm{p}}$ and $M_{\rm R}\geq M_{\rm T}$, from $\unfnot{\ca{\tilde{\ten{Y}}}}{[3,4],[1,2]}\cdot\left[\left(\bar{\ma{H}}_{\rm{p}}\krp(\ma{I}_N\otimes\ma{1}_{M_{\rm T}}^\trans)\right)^\trans\right]^\ca{\pinv}\approx \left[\bar{\ma{C}}^\prime\krp\bar{\ma{S}}\right]$ based on LSKRF, we obtain $\hat{\bar{\ma{S}}}$ and $\hat{\bar{\ma{C}}}^\prime$. However, the matrices $\hat{\bar{\ma{S}}}$ and $\hat{\bar{\ma{C}}}^\prime$ are estimated up to one complex scaling ambiguity per column. We exploit the first row of the matrix ${\bar{\ma{C}}}^\prime$ to estimate this ambiguity (recall that the elements of the first row of the matrix ${\bar{\ma{C}}}^\prime$ are set to one). After resolving the scaling ambiguity, we propose to iterate between the equations~\eqref{US}-\eqref{UH} to enhance the accuracy of the receiver. Hence, we propose two receivers RC-KR (Random Coding-Khatri-Rao) and RC-KR+ALS (Random Coding-Khatri-Rao+Alternating Least-Squares) for randomly coded MIMO-OFDM systems. These two algorithms are summarized in Algorithm~\ref{Alg:RC-KR} and Algorithm~\ref{Alg:RC-KR+ALS}, respectively. The RC-KR receiver exploits the LSKRF to compute an estimate of the symbol matrices $\bar{\ma{S}}$ and $\bar{\ma{C}}^\prime$, assuming that $M_{\rm R}\geq M_{\rm T}$, the first row on the matrix $\bar{\ma{C}}^\prime$ contains only ones, and a pilot based channel estimate $\bar{\ma{H}}_{\rm{p}}$ is already available. The initial steps of the RC-KR+ALS receiver are equivalent to the RC-KR receiver. In a following steps using the RC-KR+ALS receiver, we estimate the channel matrix and both symbol matrices in an ALS fashion. Therefore, the RC-KR+ALS receiver exploits \ccb{the} LSKRF to initialize the ALS algorithm. The ALS algorithm is stopped if it exceeds the \ccb{maximum} number of iterations that is set to 5, reaches a predefined minimum of the cost function $\honorm{\tilde{\ten{Y}} - {\ten{I}_{4,M_{\rm T}N}}\times_1 (\ma{I}_N\otimes\ma{1}_{M_{\rm T}}^\trans) \times_2 \hat{\bar{\ma{H}}} \times_3  \hat{\bar{\ma{S}}}\times_4  \hat{\bar{\ma{C}}}^\prime}^2/\honorm{\tilde{\ten{Y}}}^2,$ or if the error of the cost function has not changed within two consecutive iteration\cca{s}. \cca{The RC-KR algorithm has \ccb{a} higher computational complexity than the RC-KR+ALS algorithm due to the additional ALS iterations\ccb{, as shown in Algorithm~7.}} 

%% file: SimulationResults.tex
In this section, we evaluate the performance of the proposed receivers for MIMO-OFDM systems using Monte-Carlo simulations. First, we compare the performance of ZF, ILSP, and RLSP, (i.e., Algorithms~\ref{ZF}-\ref{RLSP}) using {5000 realizations}. We consider a $2 \times 2$ OFDM system, with $K$ frames, and $N= 128$ subcarriers. The pilot symbols are transmitted on every third subcarrier such that $\Delta F = 3$ and only during the first frame, i.e, $\Delta K = K$. Using these pilots\cca{,} we obtain a pilot based channel estimate with which we initialize all of the algorithms. The transmitted data symbols are independent and modulated using 4-QAM (Quadrature Amplitude Modulation). The frequency selective propagation channel is modeled according to the 3GPP (3rd Generation Partnership Project) Pedestrian A channel (Ped A)~{\cite{ITU97}}. The duration of the cyclic prefix is 32 samples and the weighting factor $\alpha = 1${, for the recursive LS. The maximum number of iterations for the iterative algorithm is set to 7}.

\begin{figure}[!t]
	\centering
	\includegraphics[width=.8\columnwidth]{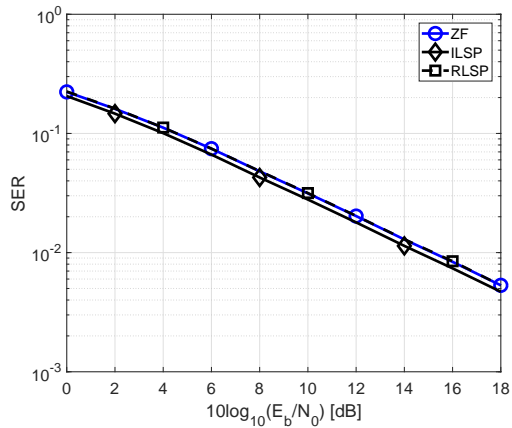}
	\caption{SER for a $2\times2$ OFDM system with parameters $\Delta K = 8, \Delta F = 3, N = 128, K = 8.$ }
	\label{fig:K8}
\end{figure}

In Fig.~\ref{fig:K8} we depict the SER (Symbol Error Rate) as a function of the $E_{\rm b}/N_0$ (energy per bit/ noise power spectral density) in dB for $K=8$.
The performance of ILSP and RLSP is similar to the ZF performance and it depends on the number of frames. As shown in Fig.~\ref{fig:K8}, increasing the number of frames leads to a slightly better SER than {using a} ZF receiver. Note that the transmitted data symbols are independent and randomly drawn with no guarantee that the matrices $\tilde{\ma{S}}_n$ are of rank $M_{\rm T}$. Therefore, in many cases the number of iterations is equal to one. In all of the simulated cases the iterative algorithm converges after 3 iterations. As in \cite{TVP96}, we also observe that the iterative algorithms have a better performance than the recursive one for an increased number of frames. However, the recursive algorithm, RSLP requires less computational complexity than the iterative one, ILSP.

Next, we compare the performance of the traditional frequency domain ZF receiver, the proposed Khatri-Rao (KR) receiver (see Algorithm~\ref{Alg:KR}) and the proposed Khatri-Rao receiver with one additional LS iteration (see Algorithm~\ref{Alg:KR+LS}). {In \ccb{these} simulations, \ccb{we average the results over} 5000 realizations and Ped A channel \cite{ITU97} are \ccb{assumed}.}
\begin{figure}[!t]
	\centering
	\includegraphics[width=.8\columnwidth]{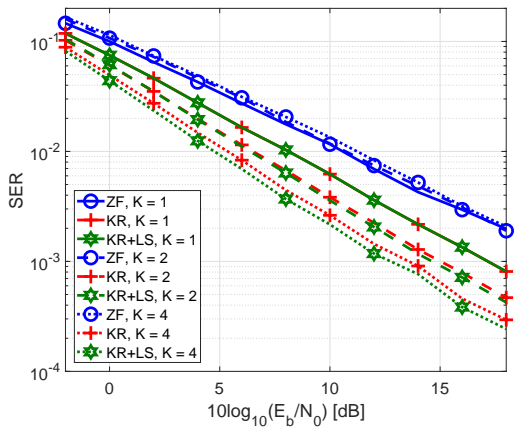}
    \vspace{-2ex}
	\caption{SER comparison for different numbers of transmitted blocks.}
	\label{fig:SERComparisonK}
\end{figure}
In Fig.~\ref{fig:SERComparisonK}, we depict the SER is as a function of $E_{\rm b}/N_0$ for different numbers of transmitted blocks. \ccb{In this case}, we consider a MIMO system with the following parameters $N=128$, $Q=2$, $M_{\rm T} =2$, $M_{\rm R}=2$, $\Delta K = K$, $\Delta F = 4$ and different numbers of blocks $K$ (the number of blocks is indicated in the legend). Note that the KR and the KR+LS receivers benefit from the increased number of frames as the channel has been kept constant during the $P=Q\cdot K$ frames. Moreover, as the number of frames {increases}{,} the advantages of the enhancement via LS become more pronounced.

\begin{figure}[!t]
	\centering
	\includegraphics[width=.8\columnwidth]{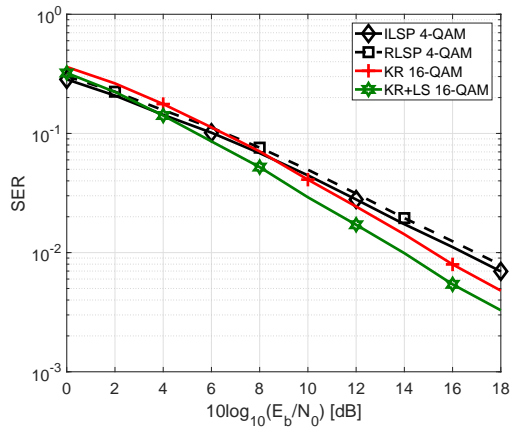}
	\caption{SER for $2\times2$ OFDM and KR coded OFDM systems, $N=128$ and $\Delta F = 10$. The OFDM system has the following parameters $K=10$, $\Delta K = 10$, and the symbols are modulated using 4-QAM. The KR coded OFDM system has the following parameters $K=5$, $\Delta K = 5$, $Q=2$, $P=KQ=10$ and the symbols are modulated using 16-QAM. Hence, both systems  transmit 2 \ccb{bits/symbol}.}
	\label{fig:SER_KRvsOFDM_16QAM_4QAM}
\end{figure}
\cca{In another experiment}, we compare the receivers proposed in Section~\ref{OFDM} (ILSP and RLSP) for a MIMO-OFDM system with the receivers proposed in Section~\ref{KR-OFDM} for a Khatri-Rao coded MIMO-OFDM system. We assume that both systems have $N=128$ subcarriers, ${M_\rm T}=2$ transmit antennas, \ccb{and} ${M_\rm R}=2$ receive antennas. Moreover, for both systems we assume that $\Delta F=10$ is the subcarrier spacing between two pilot symbols in the frequency domain and $\Delta K=K$ is {the} spacing between two pilot symbols in the time domain. The OFDM system has the following parameters $K=10$, $\Delta K = 10$, and the symbols are modulated using 4-QAM. The KR coded OFDM system has the following parameters $K=5$, $\Delta K = 5$, $Q=2$, $P=KQ=10$ and the symbols are modulated using 16-QAM. Hence, we transmit 2 \ccb{bits/symbol} with both systems. In Fig.~\ref{fig:SER_KRvsOFDM_16QAM_4QAM}, we depict the SERs for these two systems. The KR receiver has similar accuracy to  the ILSP and the RLSP algorithms that improves with the increased SNR. The KR+LS receiver outperforms the ILSP algorithm and the KR algorithm in terms of SER. Recall that the KR coded OFDM model in equation~\eqref{RXSignal} has \cca{a richer} tensor structure than the OFDM model in equation~\eqref{RXCP} due to the coding. The KR algorithm and the KR-LS algorithm \cca{effectively} exploit this structure to estimate the channel and the symbols. Note that the KR-LS algorithm computes an improved estimate of the scaling matrix. Therefore, KR-LS leads to lower SER \cca{levels} than the ILSP and KR \cca{algorithms}.

\cca{In our next experiment}, we evaluate the performance of the proposed RC-KR and RC-KR+ALS receivers for randomly coded MIMO-OFDM systems using Monte-Carlo simulations. We consider $2\times 2$ systems, with $N=128$ subcarriers, \ccb{and} $P = KQ$ frames. Moreover, the spacing between two pilot symbols in the time domain and in the frequency domain \cca{are} denoted by $\Delta K$ and $\Delta F$ respectively. The frequency selective propagation channel is modeled according to the {3GPP} Pedestrian A channel~\cite{ITU97}. The duration of the cyclic prefix is 32 samples. {In \cca{these} simulations we \cca{consider} 5000 realizations.}
\begin{figure}[!t]
	\centering
	\includegraphics[width=.8\columnwidth]{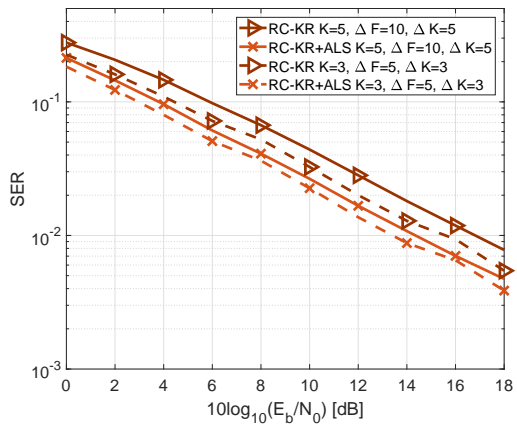}
	\caption{SER for {a} $2\times2$ randomly coded OFDM system with parameters $N=128$, $Q=2$, $K$, $\Delta K$, $\Delta F$, and the symbols are modulated using 4-QAM. The parameters $K$, $\Delta K$ and $\Delta F$ are indicated in the legend.} 
	\label{fig:NumofPilots}
\end{figure}
In Fig.~\ref{fig:NumofPilots}, we provide an SER comparison for two scenarios. For both scenarios, we assume $2\times 2$ randomly coded OFDM system, $Q=2$, and the symbols are modulated using 4-QAM modulation. Moreover, $K=5$, $\Delta F=10$, and $\Delta K= 5$, for the first scenario, whereas for the second scenario $K=3$, $\Delta F=5$, and $\Delta K= 3$. Hence, in the first scenario we estimate more symbols than in the {second} scenario, using less pilot symbols. As expected, we achieve {a} lower SER if more pilot symbols are used because they lead to {a} more accurate initial pilot based channel estimate. Moreover, in Fig.~\ref{fig:NumofPilots} we see that the RC-KR+ALS receiver outperforms the RC-KR receiver. Thus, we benefit from the additional iterations and {from} exploiting the complete tensor structure. In contrast to RC-KR, RC-KR+ALS also estimates the channel matrix. Furthermore, the accuracy gain of the RC-KR+ALS receiver is more pronounced if we initialize the RC-KR+ALS with {a} less accurate pilot based channel estimate (the gain is more pronounced for the solid lines than for the dashed lines in Fig.~\ref{fig:NumofPilots}).

\begin{figure}[!t]
	\centering
	\includegraphics[width=.8\columnwidth]{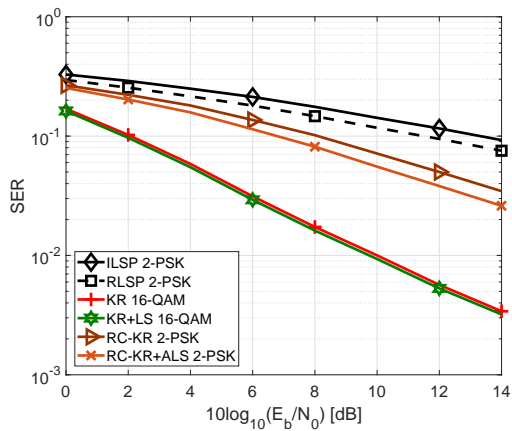}
	\caption{\ccc{SER for $4\times4$ KR coded OFDM, randomly coded OFDM, and traditional OFDM systems for $N=128$ and $\Delta F = 10$. The KR coded OFDM system \ccd{assumes} $K=2$, $\Delta K = 2$, $Q=4$, $P=KQ=8$ and the symbols are modulated using 16-QAM. The randomly coded OFDM system \ccd{assumes} $K=2$, $\Delta K = 2$, $Q=4$, $P=KQ=8$ and the symbols are modulated using 2-PSK. The OFDM system has the following parameters $K=8$, $\Delta K = 8$, and the symbols are modulated using 2-PSK.}}
	\label{fig:KRCoded-RCKR_16QAMvs2QAM}
\end{figure}
\ccc{Finally, in Fig.~\ref{fig:KRCoded-RCKR_16QAMvs2QAM}, we depict the SER {performance} for a $4\times4$ MIMO system, \ccd{considering the following receivers: (i) ILSP receiver (Algorithm~\ref{ILSP}), (ii) RLSP receiver (Algorithm~\ref{RLSP}), (iii) KR receiver (Algorithm~\ref{Alg:KR}), (iv) KR-LS receiver (Algorithm~\ref{Alg:KR+LS}), (v) RC-KR receiver (Algorithm~\ref{Alg:RC-KR}), and (vi) RC-KR+ALS (Algorithm~\ref{Alg:RC-KR+ALS}).} \ccd{To} {ensure} a fair comparison \ccd{in terms of spectral efficiency, the following parameters were chosen for the different receivers:} The KR coded OFDM system \ccd{assumes} $N=128$, $\Delta F = 10$, $K=2$, $\Delta K = 2$, $Q=4$, $P=KQ=8$ and the symbols are modulated using 16-QAM. \ccd{For} the RC coded OFDM system \cce{we assume} $N=128$, $\Delta F = 10$, $K=2$, $\Delta K = 2$, $Q=4$, $P=KQ=8$ and the symbols are modulated using \ccd{BPSK symbols}. The OFDM system \ccd{assumes} $N=128$, $\Delta F = 10$, $K=8$, $\Delta K = 8$, and \ccd{BPSK symbols}. 
We see that the RC-KR receiver ourperforms ILSP and RLSP algorithms. In addition, the KR and KR-LS receivers for KR-coded OFDM have different slopes than the uncoded {OFDM} and the randomly coded OFDM, exhibiting a better performance, as expected. } 

%% file: ChannelTensor.tex
\section*{Derivation of the 4-way channel tensor in the frequency domain and its unfoldings} \label{App:Channel Tensor}
Let us assume a MIMO-OFDM system with $M_{\rm T}$ transmit antennas and $M_{\rm R}$ receive antennas. Such a system is depicted in Fig.~\ref{fig:MIMOSystem}, for $M_{\rm T}=2$ and $M_{\rm R}=3$. As shown in Section~\ref{OFDM}, we can define a 4-way channel tensor $\tilde{\ten{H}}\in\compl^{N \times N\times M_{\rm R} \times M_{\rm T}}$ in equation~\eqref{ChannelTensor} by concatenating the channel tensors for each transmit antenna, i.e., $\tilde{\ten{H}}_{\rm R}^{(m_{\rm T})}\in\compl^{N \times N\times M_{\rm R}}$ along the 4-mode. The tensors $\tilde{\ten{H}}_{\rm R}^{(m_{\rm T})}\in\compl^{N \times N\times M_{\rm R}}$ contain the channel vectors for the $m_{\rm T}$-th transmit antenna and all receive antennas as defined in equation~\eqref{eq:ChannelTensorPerTX}, for $m_{\rm T}=1,\ldots M_{\rm T}$. Recall that these tensors have a CP structure, i.e. $\tilde{\ten{H}}_{\rm R}^{(m_{\rm T})} =\ten{I}_{3,N}\times_3\tilde{\ma{H}}_{\rm R}^{(m_{\rm T})}$, for $m_{\rm T}=1,\ldots M_{\rm T}$. The matrices $\tilde{\ma{H}}_{\rm R}^{(m_{\rm T})}$ ($m_{\rm T}=1,\ldots M_{\rm T}$) are defined in equation~\eqref{eq:ChannelTensorPerTXCP}. Hence, the 4-way channel tensor is
\begin{align*}
\tilde{\ten{H} } 
&=\begin{bmatrix}
\tilde{\ten{H}}_{\rm R}^{(1)} & \sqcup_4 & \tilde{\ten{H}}_{\rm R}^{(2)} &  \sqcup_4 & \ldots & \tilde{\ten{H}}_{\rm R}^{(M_{\rm T})}
\end{bmatrix}
\end{align*}
We can rewrite this concatenation by means of an outer product with a pining vector $\ma{e}_{m_{\rm T}}$. Moreover, if we substitute the CP structure of the tensor $\tilde{\ten{H}}_{\rm R}^{(m_{\rm T})}$, we get
\begin{align*}
\tilde{\ten{H} }  &= \sum_{m_{\rm T}=1}^{M_{\rm T}} \tilde{\ten{H}}_{\rm R}^{(m_{\rm T})} \circ \ma{e}_{m_{\rm T}} \\&= \sum_{m_{\rm T}=1}^{M_{\rm T}} \left(\ten{I}_{3,N}\times_1\ma{I}_N\times_2\ma{I}_N\times_3\tilde{\ma{H}}_{\rm R}^{(m_{\rm T})}\right)\circ \ma{e}_{m_{\rm T}}.
\end{align*}
\ccb{Replacing} the outer product by an $n$-mode product, we have
\begin{align}
\tilde{\ten{H} }  =\sum_{m_{\rm T}=1}^{M_{\rm T}} \ten{D}\times_1\ma{I}_N\times_2\ma{I}_N\times_3\tilde{\ma{H}}_{\rm R}^{(m_{\rm T})}\times_4 \ma{e}_{m_{\rm T}}, \label{eq:BTDofChannelTensor}
\end{align}
where $\ten{D}_{(.,.,.,1)}=\ten{I}_{3,N}$. Note that the tensor $\ten{D}\in \real^{N\times N\times N\times 1}$ is a 4-way tensor, but its 4-mode is a singleton dimension. We can define this tensor in terms of a Kronecker product, which yields $\ten{D} = \ten{I}_{4,1} \otimes \ten{I}_{3,N}$. Equation~\eqref{eq:BTDofChannelTensor} represents a very special BTD where the block terms are equivalent in all modes, but the 3-mode and the 4-mode. Next, we can replace the sum in~\eqref{eq:BTDofChannelTensor} with a block diagonal core tensor and factor matrices partitioned accordingly.
\begin{align*}
\tilde{\ten{H} }  = &\blockdiagof{ \ten{I}_{4,1} \otimes \ten{I}_{3,N}}_{m_{\rm T}=1}^{M_{\rm T}}\times_1\begin{bmatrix}
{\ma{I}_N} & \ldots & {\ma{I}_N}
\end{bmatrix}\\&\times_2\begin{bmatrix}
{\ma{I}_N} & \ldots & {\ma{I}_N}
\end{bmatrix} \times_3 \underbrace{\begin{bmatrix}
\tilde{\ma{H}}_{\rm R}^{(1)} & \ldots & \tilde{\ma{H}}_{\rm R}^{(M_{\rm T})} \end{bmatrix}}_{\tilde{\ma{H}}}\\&\times_4\begin{bmatrix}
{\ma{e}_1} & \ldots & {\ma{e}_{M_{\rm T}}}
\end{bmatrix}
\end{align*}
Further, we rewrite the block diagonal structure and the partitioned factor matrices using Kronecker products
\begin{align}
\tilde{\ten{H}} =&(\ten{I}_{4,M_{\rm T}}\otimes\ten{I}_{3,N}) \times_1 (\ma{1}_{M_{\rm T}}^\trans\otimes\ma{I}_{N}) \times_2 (\ma{1}_{M_{\rm T}}^\trans\otimes\ma{I}_{N}) \\&\times_3 \tilde{\ma{H}} \times_4 \ma{I}_{M_{\rm T}}. \label{ChannelTensorExplicitStructure}
\end{align}
This last equation explicitly reveals the structure of the channel tensor $\tilde{\ten{H}}$. Exploiting this structure, we can define any of the tensor unfoldings. For the generalized unfolding $\unfnot{\tilde{\ten{H}}}{[1,3],[2,4]}$, from equation~\eqref{ChannelTensorExplicitStructure}, we get
\begin{align}
\unfnot{\tilde{\ten{H}}}{[1,3],[2,4]} = &\left[\tilde{\ma{H}}\kron(\ma{1}_{M_{\rm T}}^\trans\otimes\ma{I}_{N})\right] \notag
\\&{\unfnot{\ten{I}_{4,M_{\rm T}}\otimes\ten{I}_{3,N}}{[1,3],[2,4]}\left[\ma{I}_{M_{\rm T}} \kron \ma{1}_{M_{\rm T}}^\trans\otimes\ma{I}_{N} \right]} \label{eq:unfg}
\end{align}
Next, we have
\begin{align*}
\unfnot{\ten{I}_{4,M_{\rm T}}\otimes\ten{I}_{3,N}}{[1,3],[2,4]}\left[\ma{I}_{M_{\rm T}} \kron \ma{1}_{M_{\rm T}}^\trans\otimes\ma{I}_{N} \right] = \ma{I}_{N{M_{\rm T}}} \krp \ma{I}_{N M_{\rm T}}
\end{align*}
for the second part in \eqref{eq:unfg}. Recognize that {$\ma{I}_{N{M_{\rm T}}} \krp \ma{I}_{NM_{\rm T}}=\ma{J}_{NM_{\rm T}}$} is a selection matrix that converts a Kronecker product into a Khatri-Rao. Using this property, \eqref{eq:unfg} becomes
\begin{align}
\unfnot{\tilde{\ten{H}}}{[1,3],[2,4]} = \tilde{\ma{H}}\krp(\ma{1}_{M_{\rm T}}^\trans\otimes\ma{I}_{N}). \label{eq:ChannelTensor13-24}
\end{align}

Moreover, the generalized unfolding $\unfnot{\tilde{\ten{H}}}{[1,3],[4,2]}$ can also be derived directly from equation~\eqref{ChannelTensorExplicitStructure}. However, to simplify the final result is not \clh{straightforward} because $N$ is the faster rising index along the columns of the factor matrix $\tilde{\ma{H}}$ in equation~\eqref{ChannelTensorExplicitStructure}. On the other hand, $M_{\rm T}$ varies faster than $N$ along the columns in the generalized unfolding $\unfnot{\tilde{\ten{H}}}{[1,3],[4,2]}$. Therefore, we derive this generalized unfolding by means of a permutation matrix $\ma{P} \in \real^{NM_{\rm T} \times M_{\rm T}N}$. The permutation matrix $\ma{P}$ reorders the columns such that the faster increasing index is $M_{\rm T}$ instead of $N$ and is defined as $[\tilde{\ten{H}}]_{([1,3],[4,2])}=[\tilde{\ten{H}}]_{([1,3],[2,4])}\cdot\ma{P}$. Hence, \begin{align}
[\tilde{\ten{H}}]_{([1,3],[4,2])}= \left[\tilde{\ma{H}}\krp(\ma{1}_{M_{\rm T}}^\trans\otimes\ma{I}_{N})\right]\cdot\ma{P}. \label{nexkoy}
\end{align}
Considering that the permutation matrix $\ma{P}$ reorders the columns in equation~\clh{\eqref{nexkoy}} and the Khatri-Rao product is a column-wise operator (Khatri-Rao product is column-wise Kronecker product)\clh{,} the following equality holds
\begin{align*}
[\tilde{\ten{H}}]_{([1,3],[4,2])} &= \left[\tilde{\ma{H}}\krp(\ma{1}_{M_{\rm T}}^\trans\otimes\ma{I}_{N})\right]\cdot\ma{P} \\&= \left[\tilde{\ma{H}}\cdot\ma{P}\right]\krp\left[(\ma{1}_{M_{\rm T}}^\trans\otimes\ma{I}_{N})\cdot\ma{P}\right].
\end{align*}
\clh{The permutation matrix for $M_{\rm T}=2$ and $N=3$ is given by}
\begin{align*}
\ma{P} = \clh{\begin{bmatrix}
1&	0&	0&	0&	0&	0 \\
0&	0&	0&	1&	0&	0 \\
0&	1&	0&	0&	0&	0 \\
0&	0&	0&	0&	1&	0 \\
0&	0&	1&	0&	0&	0 \\
0&	0&	0&	0&	0&	1
\end{bmatrix}}\clh{.}
\end{align*}
Finally, using $(\ma{1}_{M_{\rm T}}^\trans\otimes\ma{I}_{N})\cdot\ma{P}= (\ma{I}_{N}\otimes\ma{1}_{M_{\rm T}}^\trans)$ and defining $\bar{\ma{H}}=\tilde{\ma{H}}\cdot\ma{P}$, we get
\begin{align*}
[\tilde{\ten{H}}]_{([1,3],[4,2])} = \bar{\ma{H}}\krp(\ma{I}_{N}\otimes\ma{1}_{M_{\rm T}}^\trans). \label{eq:ChannelTensor13-42}
\end{align*} 